\documentclass[aps,prl,twocolumn,floats,epsfig,pdflatex]{revtex4-2}

\usepackage{color, ulem, verbatim, float}
\usepackage{amssymb,amsbsy,amsmath,mathrsfs}
\usepackage{amssymb}
\usepackage{amsbsy}
\usepackage{amsmath}
\usepackage{float}
\usepackage{graphicx}
\usepackage{xcolor}
\usepackage{bm}
\usepackage{mathtools} 
\usepackage{placeins} 
\usepackage{natbib}
\definecolor{bostonuniversityred}{rgb}{0.8, 0.0, 0.0}
\definecolor{bittersweet}{rgb}{1.0, 0.44, 0.37}
\definecolor{MyTealBlue}{RGB}{0, 174, 179}
\definecolor{ForestGreen}{RGB}{34,139,34}
\definecolor{Mulberry}{RGB}{184, 12, 227}
\definecolor{Rblue}{RGB}{65, 105, 225} 
\usepackage[plainpages=false,pdfpagelabels,colorlinks=true,linkcolor=Mulberry,urlcolor=blue,citecolor=bostonuniversityred,pdftitle={},pdfauthor={},pdfdisplaydoctitle={},pdfduplex=DuplexFlipLongEdge]{hyperref}

\newcommand\bea{\begin{eqnarray}}
\newcommand\eea{\end{eqnarray}}
\newcommand\beq{\begin{equation}}
\newcommand\eeq{\end{equation}}

\def\red{\textcolor{red}}

\usepackage{parskip}

\begin{document}
\title{Engineering Ergodicity-Breaking and Stable Emergent Conservation Laws in a \\ Generic Quantum Many-body System Under Strong Periodic Drive}
\author{Sagnik Chaudhuri$^1$, Anirban Das$^1$, Asmi Haldar$^{2}$, and
Arnab Das$^1$}

\affiliation{$^1$ Indian Association for the Cultivation of Science, School of Physical Sciences,\\ 2A \& 2B Raja S. C. Mullick Road, Kolkata 700032, West Bengal, India \\
$^2$ Paul Sabatier University, CNRS, Laboratoire de Physique Th\'eorique, Toulouse,  France}

\begin{abstract}
A periodically driven quantum many-body system is believed to approach a locally infinite-temperature-like state as dictated by the ergodicity hypothesis, when all the couplings/parameters are of the same order. When the drive is stronger than a threshold value, a large class of systems is shown to break this mandate, exhibiting Dynamical Freezing (DF): ergodicity is broken due to the emergence of local operators with permanent memory of their initial values~\href{https://journals.aps.org/prx/abstract/10.1103/PhysRevX.11.021008}{(PRX {\bf 11}, 021008, 2021)}. However, an exception to DF has recently been noted in a particular setting~\href{https://journals.aps.org/prx/abstract/10.1103/4w5w-57my}{(PRX {\bf 16}, 011041, 2026)}. Here we identify the condition for ergodicity-breaking in a system under strong drive, and show how to stabilize DF by {\it exact, local} engineering. We further show that the Floquet Hamiltonian $H_{eff} = (i/T)\ln{[U(T)]}$ is non-local when ordered by spin-flips, in a non-DF system, while it turns {\it quasi-local} under our simple, local DF engineering that keeps the Hamiltonian $H(t)$ still non-integrable at each $t$. This perplexing non-universality is absent in equilibrium, where non-integrability is believed to imply ergodicity and thermalization universally. Our work paves a way to engineering novel ergodicity-broken interacting Floquet phases of matter in various quantum platforms, and also endows DF-assisted quantum error mitigation and state stabilization techniques (see~\href{https://www.nature.com/articles/s41586-026-10585-6}{Nature {\bf 653}, 1027 (2026)}) with unprecedented versatility and flexibility. 
\end{abstract}

\maketitle
It is widely expected that a generic closed quantum system described by a clean, non-integrable, time-independent Hamiltonian satisfies the eigenstate thermalization hypothesis (ETH)~\cite{Srednicki, Rigol_Nature}. 
The scenario is far richer when such a system is driven by making its Hamiltonian time-dependent. For example, if such a system is subjected to a ``generic'' periodic drive (not fine-tuned to trivially cancel the effect of driving), the common, ergodicity-driven expectation is that it should keep absorbing energy until it reaches an infinite-temperature-like state locally~\cite{Alessio_Rigol_Floquet_ETH, LDM_Floquet_ETH}. This intuition, however, works universally only if all the system parameters are of the same order. In particular, it can break down when the drive is stronger than a system size ($L$)-independent threshold, giving rise to Dynamical Freezing ({\bf DF}, see, e.g.,~\cite{AD-DMF, Asmi_DF_PRX_2021, DF_TDL, Lu-Dong_DF_Nature, AD-SDG, Mahesh_Freezing, Onset,   Asmi_DF_Krylov,  Naveen_Dynamical_Freezing, Debanjan_PRL, Debanjan_PRX_Instantons, Bhaskar_DF, Madhumita_DF_Exp, Asmi_Flq_Rev}), where the drive creates certain local operators - the Emergent Conserved Operators ({\bf ECO})s, whose expectation values show small, permanently bounded fluctuations close to their initial values, with perpetually steady averages. DF thus prevents Floquet-ETH and breaks ergodicity.

However, the mechanism of DF does not simply consist of suppressing various processes by imposing large energy costs by the large field (see, e.g.~\cite{DF_TDL}), and hence, it cannot, by default, be expected to occur whenever the drive is large (see, e.g.~\cite{Debanjan_PRX_Instanton_DF}). Here we address the broad question: what is the minimal structure necessary for DF to occur, and how it can be stabilized in a strongly driven system where it is not present by default.

\begin{figure*}[htb]
\begin{center}
\includegraphics[width=0.435\linewidth]{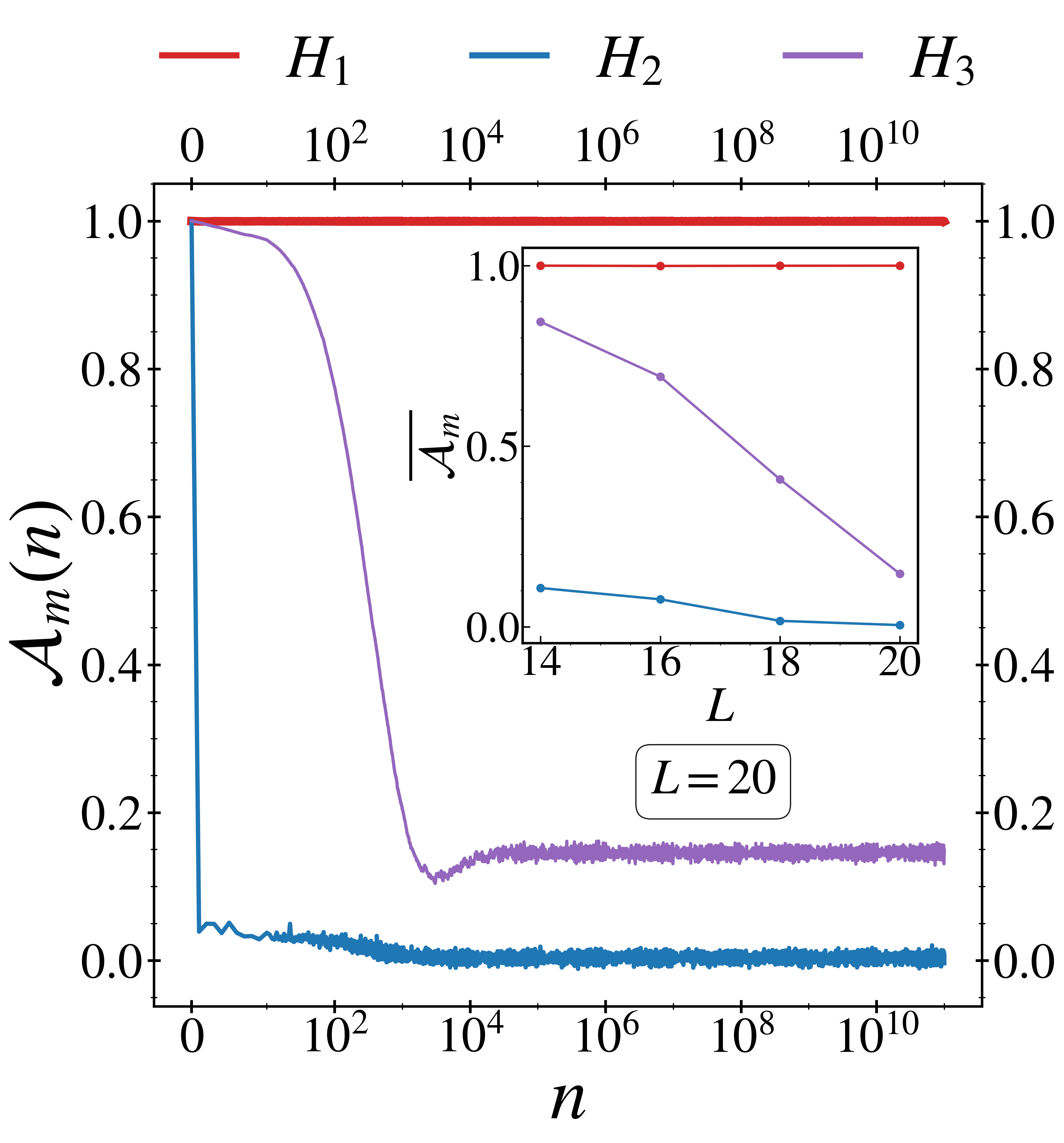}
\includegraphics[width=0.435\linewidth]{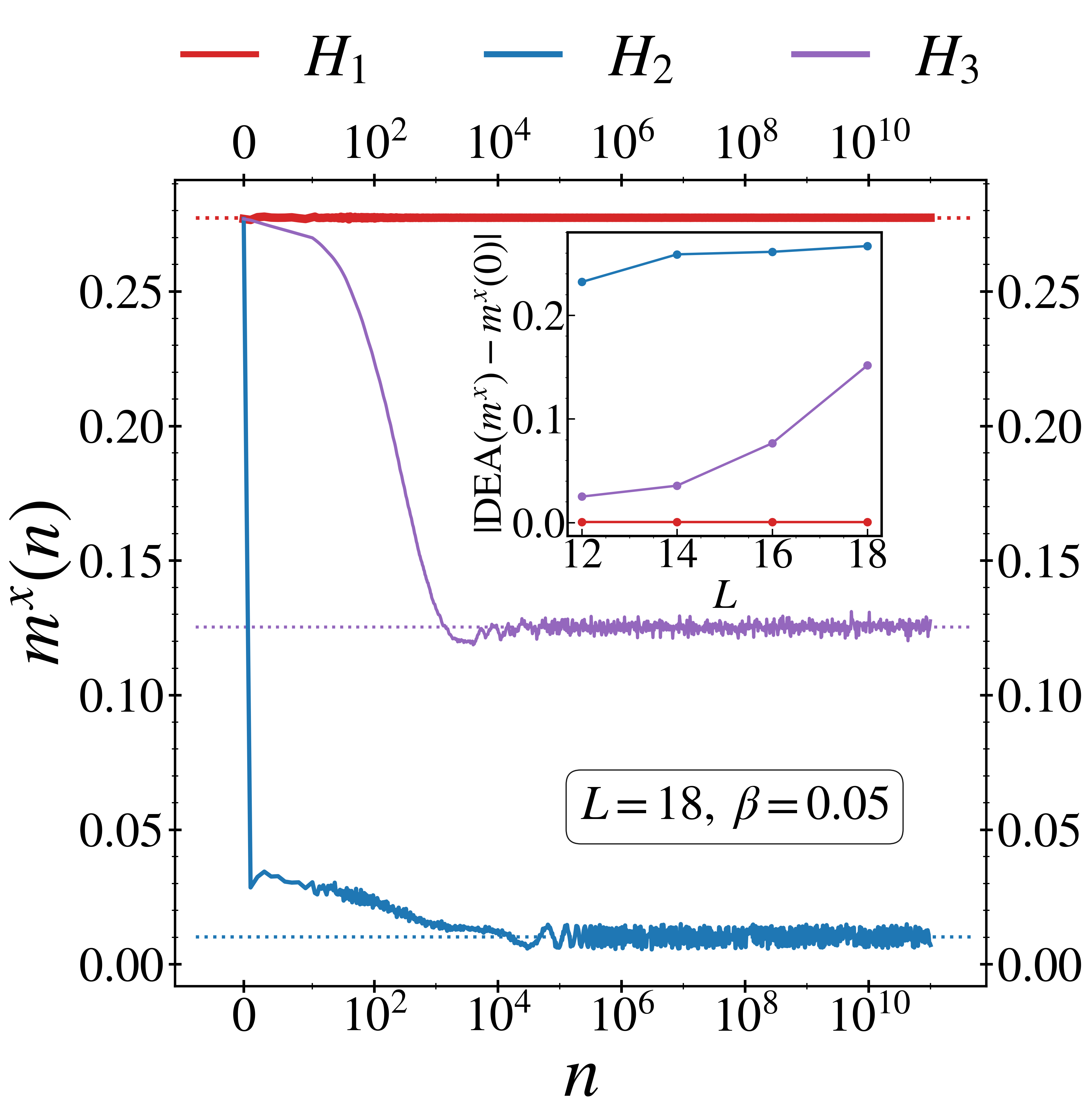}
\end{center}
\caption{{\bf Nonuniversal Floquet Thermalization and Freezing Under Strong Driving}. Left: for $L=20$, the magnetization autocorrelation remains near unity for $H_1$, decays nearly to zero for $H_2$, and reaches an intermediate plateau for $H_3$; the inset shows the late-time averaged magnetization autocorrelation $\overline{\mathcal A_m}$ for $L=14,16,18,20$ (the late-time averages have been computed over 1000 uniformly spaced cycle numbers in $10^8\leq n\leq10^{11}$). Right: starting from a finite-temperature mixed state ($\beta=0.05$), $m^x$ retains its initial value for $H_1$, while $H_2$ and $H_3$ relax to their infinite-time diagonal-ensemble averages (dotted lines); the inset shows $|\mathrm{DEA}(m^x)-m^x(0)|$ versus $L$. The decreasing plateaus of $H_2, H_3$ contrast with the retained memory of $H_1$, mirrored by the mixed-state DEA, which matches the initial magnetization only for $H_1$. For both panels, the drive amplitudes are
$h_D^x=30\omega_{_1}$ for $H_1$, $J_D=30\omega_{_1}$ for $H_2$, and $h_D^x=15\omega_{_2}$ for $H_3$. Additional correlators, amplitude scans, and details of the mixed-state preparation are given in the Supplemental Material~\cite{SM}.}
\label{fig:contrast}
\end{figure*}

\begin{figure*}[t]
\includegraphics[width=0.435\linewidth]{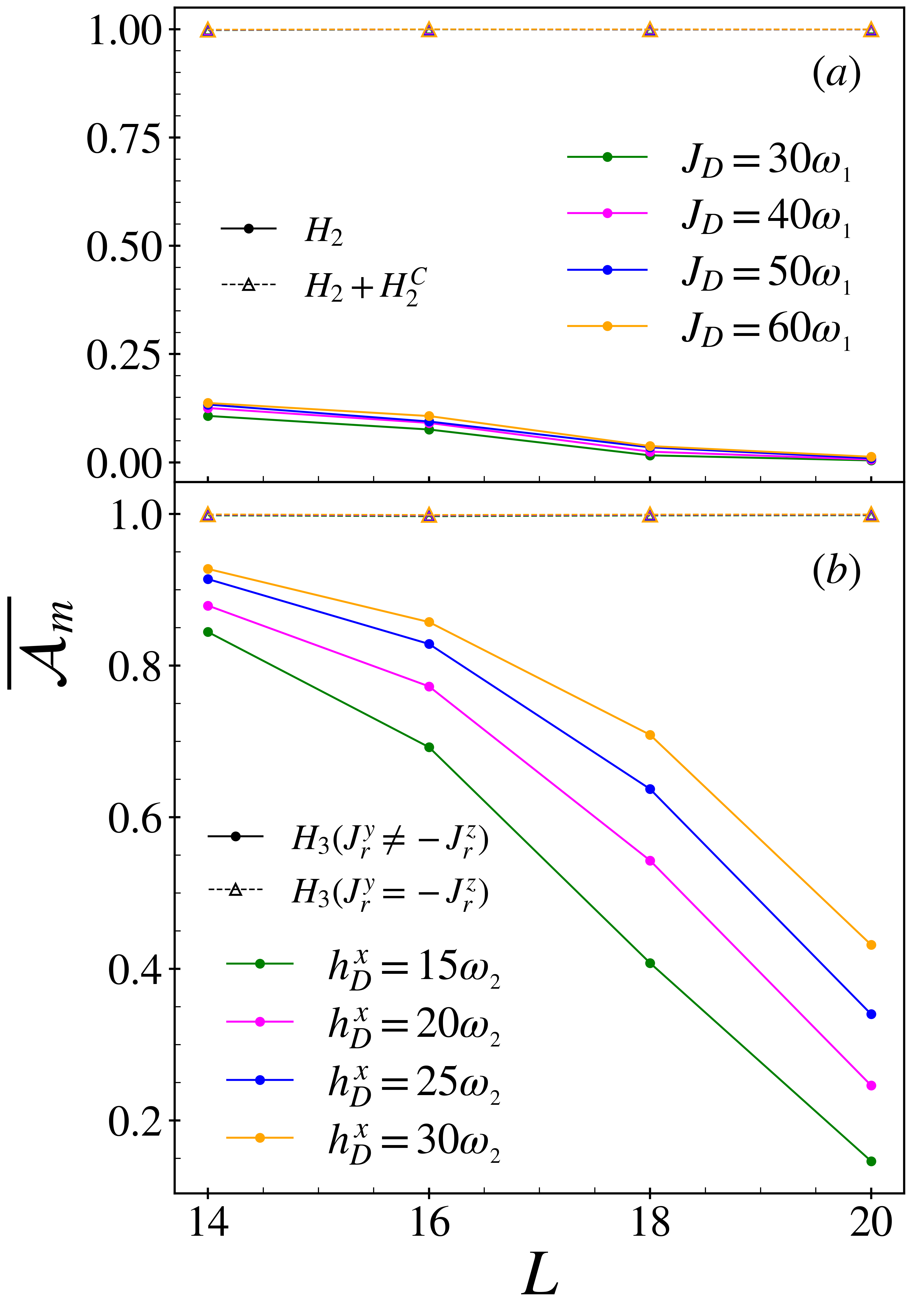}
\includegraphics[width=0.435\linewidth]{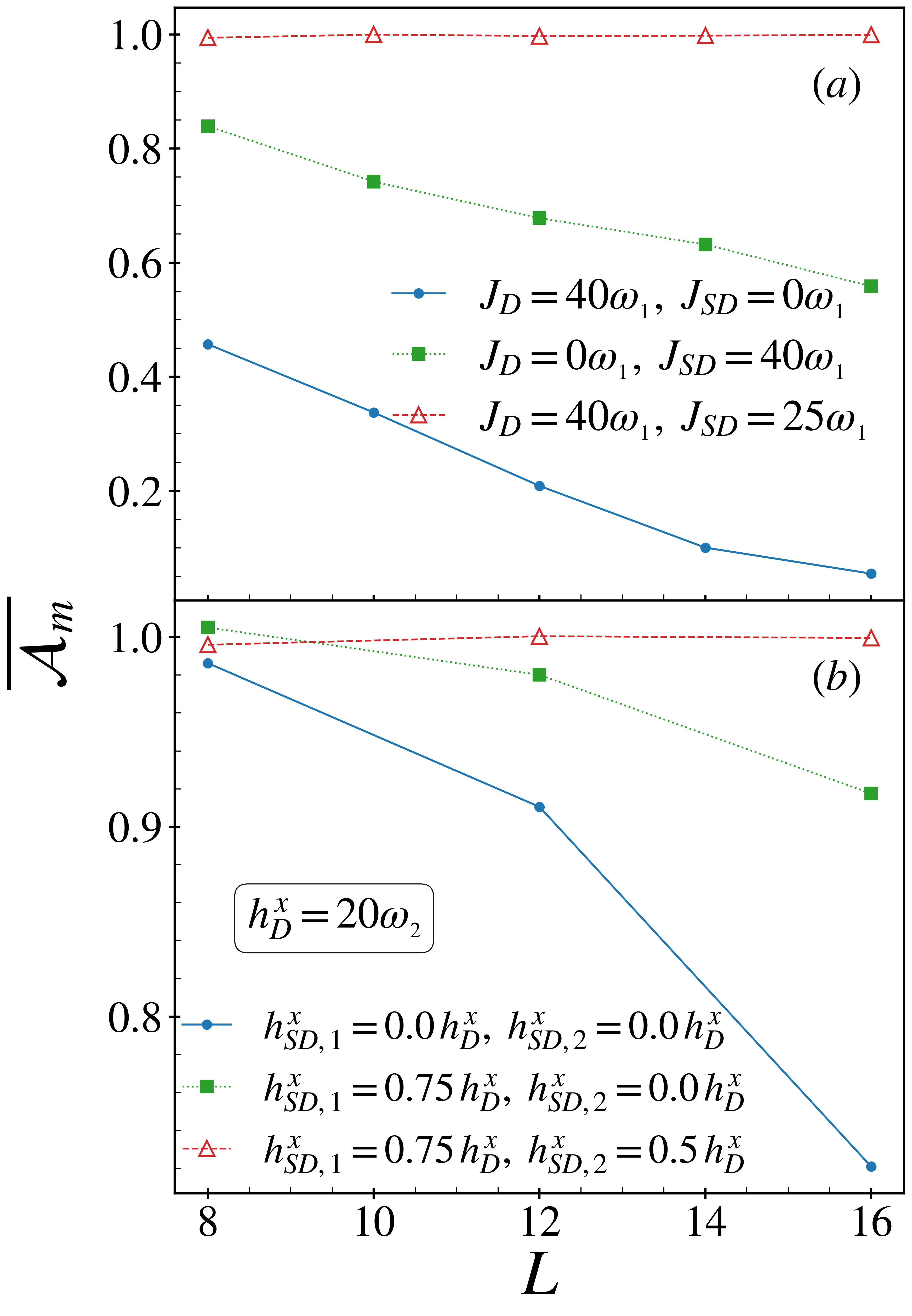}
\caption{{\bf Restoring Operator Memory by Hamiltonian Engineering:} Upper panels: $H_2$; lower panels: $H_3$. Left: static corrections give near-unity memory (triangles), whereas the original models lose memory with increasing $L$ (circles). Right: the combined drive components (red triangles) remove the downward size trend present for incomplete combinations. Amplitudes are indicated; $\omega_a$ denotes the corresponding model's frequency. The late-time averages $\overline{\mathcal A_m}$ are computed over 1000 uniformly spaced cycle numbers in $10^8\leq n\leq10^{11}$. Real-time dynamics and amplitude scans appear in the Supplemental Material~\cite{SM}.}
\label{fig:engineering}
\end{figure*}

\begin{figure}[h!]
\begin{center}
\includegraphics[width=0.95\columnwidth]{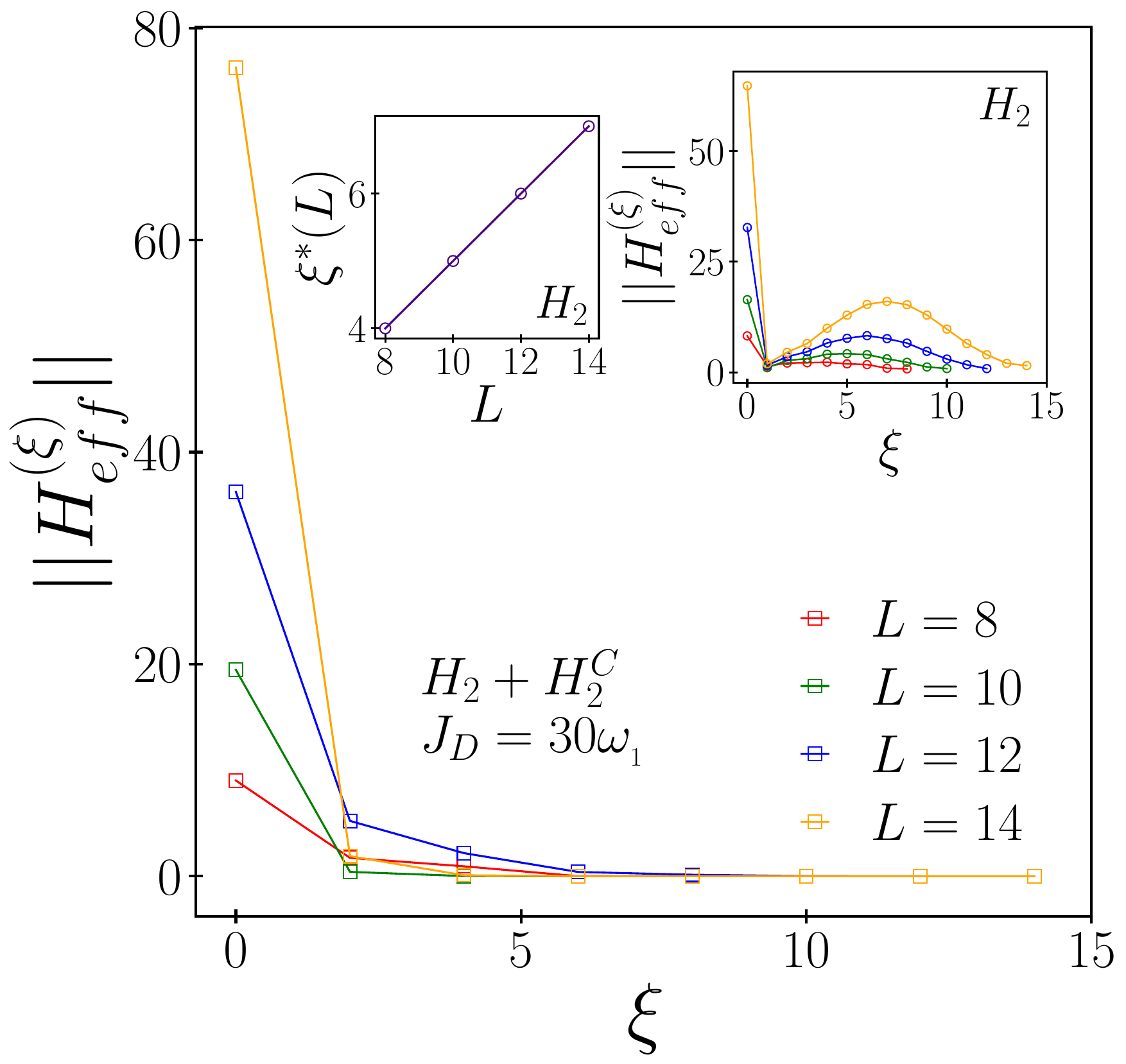}
\end{center}
\caption{{\bf Quasi-locality of $H_{eff}$ in Spin-flip Ordered Series} for the bond-driven Ising model in the strongly driven regime. The main panel
shows the Frobenius norm $\|H_{eff}^{(\xi)}\|$ of the even-$\xi$ spin-flip sectors for $H_{2}+H_{2}^{C}$ as a function of the
Hamming distance $\xi$, for $J_D=30\omega_{_1}$ and system sizes $L=8,10,12,14$. For all system sizes, the norm rapidly decays with increasing $\xi,$ and the decay sets in at approximately the same spin-flip distance, indicating a robust quasi-local structure of the
effective Hamiltonian. The right inset shows the corresponding spin-flip-resolved norm for $H_{2}$ without the correction term, where, apart from the diagonal sector $\xi=0$, the dominant weight is transferred to $\xi\simeq L/2$. The left inset shows $\xi^{*}(L)$ for $H_2$, the nonzero spin-flip order ($\xi>0$) at which the Frobenius norm $F_N(\xi)$ is maximal for a given system size $L$.    
}
\label{fig:contrast_LOE}
\end{figure}

Here, first we present contrasting cases of the presence and the absence of DF in various interacting non-integrable systems under strong driving (Fig.~\ref{fig:contrast}). Next, we present our criterion for the occurrence of DF in a generic, arbitrary, strongly driven system. Based on that, we derive the prescription for engineering DF in systems which originally do not support DF. We then numerically test the result of the engineering employed in two different ways: drive engineering and time-independent engineering (Fig.~\ref{fig:engineering}). Finally, we confirm the genuineness of the ergodicity-breaking ({\bf EB}) in a system with engineered DF by demonstrating the quasi-local character of the exact Floquet Hamiltonian $H_{eff}  = (i/T)\ln{[U(T)]}$ (Fig.~\ref{fig:contrast_LOE}), where $U(T)$ is the evolution operator over a drive period $T,$ and conclude.

%
\noindent
{\bf Occurrence and Absence of DF in Strongly Driven Systems:} Consider three different clean interacting spin chains, with
$X_i=\sigma_i^x$, $Z_i=\sigma_i^z$, and
$s_a(t)=\operatorname{sgn}[\sin(\omega_a t)]$.
The first pair has identical static $x$ parts but different driven operators:
\begin{align}
H_1(t)&=H_1^x-h^z\sum_i Z_i-h_D^x s_1(t)\sum_iX_i,
\label{eq:h1}\\
H_2(t)&=H_2^x-h^z\sum_i Z_i-J_D s_1(t)\sum_iX_iX_{i+1},
\label{eq:h2}
\end{align}
where
\begin{equation}
H_1^x=H_2^x
=-J\sum_iX_iX_{i+1}
+\kappa\sum_iX_iX_{i+2}
-h_0^x\sum_iX_i.
\label{eq:h0}
\end{equation}
The third retains a uniform field drive while changing the static interactions:
\begin{equation}
H_3(t)=H_3^x
-\sum_{r=1}^{2}J_r^z\sum_iZ_iZ_{i+r}
-h_D^x s_2(t)\sum_iX_i,
\label{eq:h3}
\end{equation}
with
$H_3^x=g^x\sum_iX_i$.
Here $T_a=2\pi/\omega_a$ and $\hbar=1$. Both frequencies remain comparable
to intrinsic energy scales; the only large parameter is the drive amplitude. Unless stated otherwise, we use the following parameter values throughout: $J=2.0$, $\kappa=0.5$, $h_0^x=0.12$, $h^z=\sqrt{3}/1.5$, and $\omega_{_1}=\phi/1.6$ for $H_1$ and $H_2$, where $\phi$ is the golden ratio, and $g^x=0.8$, $J_1^z=1.0$, $J_2^z=0.5$, and $\omega_{_2}=2\pi/3.1$ for $H_3$.

We diagnose memory using $m^x=L^{-1}\sum_iX_i$ and its normalized autocorrelation,
\begin{equation}
\mathcal A_m(n)=
\frac{\langle\psi_H|Q(nT_a)Q|\psi_H\rangle}
{\langle\psi_H|Q^2|\psi_H\rangle},\qquad Q=m^x,
\label{eq:corr}
\end{equation}
where $|\psi_H\rangle$ is Haar random and $Q(nT_a)=U_F^{-n}QU_F^n$, with $U_F$ the one-period propagator. Haar typicality connects this quantity to the infinite-temperature trace correlation. Near-unity values signal retained operator memory. 
For $H_1$ and $H_2$, the correlators $C_r^x=L^{-1}\sum_i X_iX_{i+r}$ ($r=1,2$) provide a complementary contrast beyond magnetization~\cite{SM}. \\

\noindent
{\bf Long-time Responses:} In Fig.~\ref{fig:contrast}, $\mathcal A_m(n)$ for $H_1$ stays near unity, with small fluctuations, up to $10^{11}$ cycles. For $H_2$, it rapidly loses most of its memory. Replacing the driven field with a driven interaction thus changes the response even though the static Hamiltonian is identical. The driven operator alone is also insufficient to determine the outcome. Although $H_1$ and $H_3$ both have uniform $x$-field drives, $\mathcal A_m(n)$ for $H_3$ relaxes to a plateau near $0.15$ at $L=20.$  The $ t\to\infty$ limit, i.e., the diagonal ensemble average (DEA)~\cite{Rigol_Nature} for a finite-temperature mixed-state dynamics, resoundingly agrees with these results~\cite{SM}.\\
As $L$ is increased, the inset shows the late-time averaged magnetization autocorrelation (denoted by $\overline{\mathcal A_m}$), which remains near unity for $H_1$ and decreases systematically for $H_2$ and $H_3$. Consistent with that, the mixed-state inset shows a very small separation between the initial and DEA magnetization for $H_1$ with no increase with $L,$ while an increasing separation with $L$ for $H_2,  H_3.$ Closely related clean systems thus exhibit either persistent memory or relaxation consistent with ergodicity. Neither fate follows from strong driving alone. \\

\noindent
{\bf Criteria for Stable EB:}
The comparison above suggests a simple structural criterion for stable DF.
The driven Hamiltonians considered here have the general laboratory-frame
form $H(t)=H_{stat}+H_D^x(t)$, with $H_{stat}=H^x+V$ and $H_D^x(t)=-A_D s(t)D^x$, where $A_D$ denotes the drive amplitude. All parameters in $H^x$ and $V$, together with the drive frequency $\omega$, are of $O(1)$ strength, while $A_D\gg \omega$ and all parameters in $H^x$ and $V$. Both $H^x$ and the drive operator $D^x$ are composed exclusively of mutually commuting $X$ operators, so that $[H^x,D^x]=0$, while in general $[V,D^x]\neq0$ and $[V,H^x]\neq0$.


Since $A_{D}$ is much larger than all other parameters, any series in its power, representing the dynamics, will not only diverge, but will diverge from the very onset, leaving no space even for the early orders to be asymptotic to the actual underlying function describing the true solution~\cite{Bender_Orzag_Book}. We therefore switch to the
rotating frame of the drive, with
$R(t)=e^{iA_D D^x\chi(t)}$ and
$\chi(t)=\int_0^t s(u)du$. Let $P_\gamma$ project onto the eigenspace of
$D^x$ with eigenvalue $\gamma$, and define
\begin{equation}
V_q=\sum_{\substack{\gamma,\gamma'\\ \gamma-\gamma'=q}}
P_\gamma H_{stat} P_{\gamma'} .
\end{equation}
Since $[D^x, V_q]=qV_q$, the rotating-frame Hamiltonian takes the form (for the derivation, see~\cite{SM})
\begin{equation}
H^{\mathrm{Rot}}(t)=R^\dagger(t)H_{stat}R(t)=V_0+
\sum_{q\neq0}e^{-iqA_D\chi(t)}V_q ,
\label{eq:b_rot}
\end{equation}
This decomposition separates two qualitatively different contributions. The $q\neq0$ terms rotate rapidly on a scale set by $A_D$ and their
contributions to $H_{eff}$ are suppressed by inverse powers of $A_D$, as follows, for example, from a Magnus-like expansion 
~\cite{Asmi_DF_PRX_2021}. In contrast,
$V_0=H^x+\sum_\gamma P_\gamma V P_\gamma$ is explicitly time-independent in the rotating frame, since $[D^x,V_0]=0$, and is not renormalized by the large drive. Hence, any term involving $Y$ or
$Z$ operators in $\sum_\gamma P_\gamma V P_\gamma$ survives with its bare $O(1)$ strength. $H^x$ is a linear combination of the ECOs, and any such $O(1)$ terms involving $Y$ or $Z$ operators cannot be treated as perturbations to them.

\noindent
{\bf The Rationale:} The perpetual stability of the ECO memories strongly indicates stability beyond what the inevitable $n!$-divergence ($n$ denoting the order of the expansion) of a Magnus/perturbative expansion of the Floquet Hamiltonian $H_{eff}$ na\"{\i}vely suggests, namely, a genuinely non-local $H_{eff}$  which conserves no local operator~\cite{Kuwahara_Mori_Saito_Prethermal}. But those divergences need not necessarily signify a physically divergent $H_{eff}$. Quantum electrodynamics provides a familiar example: Dyson's argument for the divergence of its coupling expansion coexists with the remarkable accuracy of perturbative predictions~\cite{Dyson_QED_Divergence_PhysRev}. Concrete QED effective-action expansions are known to be divergent but asymptotic and amenable to Borel summation~\cite{Dyson_Divergence_Resolution}. Here, in a similar spirit, we assume that the generic divergence of the expansions for $H_{eff}$ does not necessarily imply physical non-locality of $H_{eff}$ or Floquet thermalization~\cite{LDM_Floquet_ETH, Alessio_Rigol_Floquet_ETH}, and we make the following concrete conjecture. 

\textit{A Testable Conjecture:}
We conjecture the following criterion for stable EB in a DF system: $V_0$ (Eq.~\ref{eq:b_rot}) must consist exclusively of $X$ operators. All terms involving $Y$ or $Z$ operators must lie in the rapidly rotating $q\neq0$ sectors, where their effects are perturbatively suppressed by the strong drive. If this criterion is satisfied, DF is stable; otherwise, it is unstable. 
A testable consequence of this conjecture about perpetual stability of the apparently diverging perturbation series is that it {\it predicts the physical/exact $H_{eff}$ to be quasi-local} (shown later). 

\noindent
In light of this criterion, the contrast between the three models follows directly from their $q=0$ sectors:
\begin{align}
V_{0,1}&=H_1^x,\\
V_{0,2}&=H_2^x-\frac{h^z}{2}\sum_i
Z_i(1-X_{i-1}X_{i+1}),\\
V_{0,3}&=H_3^x-
\frac12\sum_{r,i}J_r^z
(Z_iZ_{i+r}+Y_iY_{i+r}),
\label{eq:b_models}
\end{align}
where $Y_i=\sigma_i^y$ and $r=1,2$. For $H_1$, $V_{0,1}$ is composed exclusively of $X$ operators. In contrast, $V_{0,2}$ and $V_{0,3}$ contain terms involving $Y$ or $Z$ operators with $O(1)$ strength. Our criterion successfully accounts for the retained operator memory in $H_1$ and its decay in $H_2$ and $H_3$. \\
The key significance of this $X$-only structure is that it renders
$V_0$ classical, i.e., constructed from mutually commuting onsite
$X$ operators and their products. 
Since both $V_0$ and $H^x$ commute with $D^x$, any nonzero projected contribution $\sum_\gamma P_\gamma V P_\gamma$ also commutes with the drive. Such an $O(1)$ contribution, involving $Y$ or $Z$ operators, cannot be treated perturbatively anymore, and the Magnus expansion for $H_{eff}$ might not even be asymptotic to an underlying convergent series.\\

\noindent
{\bf Engineering EB by Stabilizing DF}:
The stability criterion identified above suggests how to engineer stable ECOs and provides a direct test of the predictive power of our conjecture. All we have to ensure is that $V_0$ is composed exclusively of $X$ operators. This can be achieved either by modifying the time-independent part of the Hamiltonian or by modifying the drive. In the first route, a simple choice is to add the time-independent correction $\delta H_{stat}=-\sum_\gamma P_\gamma V P_\gamma$ in the laboratory frame, so that for the engineered Hamiltonian $H_{stat}'=H_{stat}+\delta H_{stat}$, one obtains $V_0=H^x$. Alternatively, the drive operator can be modified so that the unwanted $Y$ or $Z$ terms are transferred from the $q=0$ sector to the $q\neq0$ sectors, leaving $V_0=H^x$.\\
\noindent
{\bf Time-Independent Stabilization Route:} 
For $H_2$, it is convenient to resolve the global $q=0$ projection locally using $P_i^\pm=(1\pm X_{i-1}X_{i+1})/2$. A flip of spin $i$ leaves the drive eigenvalue unchanged only for oppositely aligned neighbors, so the $q=0$ component of $Z_i$ is $Z_iP_i^-$. Hence $V_{0,2}=H_2^x-h^z\sum_iZ_iP_i^-$. This unwanted static term is canceled by adding
\begin{equation}
H_2^C=h^z\sum_iZ_iP_i^-=\frac{h^z}{2}\sum_i
\bigl(Z_i-X_{i-1}Z_iX_{i+1}\bigr).
\label{eq:c_h2c}
\end{equation}
The remaining $-h^z\sum_iZ_iP_i^+$ flips spins with aligned neighbors, changing the drive energy by $\pm4J_D$; it therefore lies entirely in the rapidly rotating $q\neq0$ sectors, leaving $V_{0,2}=H_2^x$.
For $H_3$, the $q=0$ component of each $Z_iZ_{i+r}$ interaction is $\frac{1}{2}(Z_iZ_{i+r}+Y_iY_{i+r})$. Adding $+\sum_{r,i}J_r^zY_iY_{i+r}$ to $H_3$ therefore cancels this static exchange part and replaces the transverse interaction by
\begin{equation}
V_3^{\mathrm{mod}}=-\sum_{r=1}^{2}J_r^z\sum_i
\bigl(Z_iZ_{i+r}-Y_iY_{i+r}\bigr).
\label{eq:c_v3mod}
\end{equation}
The remaining $ZZ-YY$ terms create or annihilate spin pairs in the $x$ basis, changing the drive energy by $\pm4h_D^x$; they therefore lie in the rapidly rotating $q\neq0$ sectors, leaving $V_{0,3}=g^x\sum_iX_i$. Equivalently, $J_r^y=-J_r^z$ when the added interaction is written as $-J_r^yYY$, providing the same classical $V_{0,3}=H_3^x$.\\

\noindent
{\bf Stabilization by Drive Engineering:} The second route leaves $H_{stat}$ intact and adds $-s(t)\delta D$ to the laboratory Hamiltonian. For $H_2$, choose
\begin{equation}
D^x_2=\sum_i\bigl[J_D+(-1)^{i-1}J_{SD}\bigr]X_iX_{i+1}.
\label{eq:c_d2}
\end{equation}
A spin flip changes $D^x_2$ by $\pm4J_D$ for aligned neighbors and by $\pm4J_{SD}$ for opposite neighbors. With both amplitudes sufficiently large, all these processes rapidly rotate and $V_{0,2}=H_2^x$. Either component alone leaves an unsuppressed channel. 
For $H_3$, use $D^x_3=\sum_i a_iX_i$, where
\begin{equation}
\begin{aligned}
a_i&=h_D^x+h_{SD,1}^x(-1)^{i-1}+h_{SD,2}^x\eta_i,\\
\eta_i&=(-1)^{\lfloor(i-1)/2\rfloor}.
\end{aligned}
\label{eq:c_d3}
\end{equation}
Flipping a coupled pair changes the drive energy by $-2(a_ix_i+a_jx_j)$, which never vanishes if $a_i\ne\pm a_j$. Site staggering alone leaves equal fields at distance two; the additional $++--$ pattern removes this remaining channel. Choosing $h_{SD,1}^x=0.75h_D^x$ and $h_{SD,2}^x=0.5h_D^x$ gives the repeating fields $(2.25,0.75,1.25,-0.25)h_D^x$. All nearest- and next-nearest-neighbor pair flips then rapidly rotate, leaving $V_{0,3}=H_3^x$. The detailed projections are given in the Supplemental Material~\cite{SM}.\\

\noindent
\textit{Numerical Demonstration of DF Engineering:}
Figure~\ref{fig:engineering} shows the system-size dependence of the late-time averaged magnetization autocorrelation $\overline{\mathcal A_m}$ for both stabilization routes, across the displayed drive amplitudes. In each case, the engineered $H_2$ and $H_3$ retain near-unity $\overline{\mathcal A_m}$ with no systematic downward trend with increasing $L$, whereas the corresponding pristine models show a systematic decay with system size. Incomplete drive combinations likewise retain a downward size trend.Thus, both routes stabilize DF by restoring the $X$-only classical structure of $V_0$.\\

\noindent
{\bf The Underlying Quasi-Local Structure:} 
Our ECOs, exhibiting bounded fluctuations around their initial values, strongly indicate the existence of a quasi-local Floquet Hamiltonian $H_{eff}$.  A suitably truncated quasi-local $H_{eff}$ serves as an approximate but stable conservation law, and can support other such ECOs. Note that, though quasi-locality is a sufficient condition for EB it is not a necessary condition (see, e.g., ~\cite{Else_Bauer_Nayak_DTC}).   
For concreteness, we introduce the idea of an exact {\bf Spin-flip Ordered Series} for $H_{eff}$ as follows. We consider an expansion of the form
\begin{equation}
    H_{eff}
    =
    \sum_{\xi=0}^{L} H_{eff}^{(\xi)}, ~~{\rm with} ~~ 
    H_{eff}^{(\xi)}
    =
    \sum_{\substack{i,j\\
    \mathrm{dist}(i,j)=\xi}}
    H_{eff}^{i,j} ~
    |i\rangle\langle j|,
\label{Eq:LOS}
\end{equation}
\noindent where $|i\rangle, |j\rangle$ denote two $x$-bit-string states, and $dist(i,j)$ denotes the Hamming distance between them. $H_{eff}^{(\xi)}$ contains all matrix elements of
$H_{eff}$ that connect two $x$-bit-string configurations separated by
a Hamming distance $\xi.$ Thus, $\xi$ is the $x$-bit-flip order, and $\xi=0$ denotes the terms diagonal in $x$-bit-string basis ($x$-basis). We quantify the weight of each such sector
by its Frobenius norm $F_{N}:$
\begin{equation}
F_{N}^{2}(\xi)  =  \left\|H_{eff}^{(\xi)}\right\|^2
    =
    \sum_{\substack{i,j\\
    \mathrm{dist}(i,j)=\xi}}
    \left|H_{eff}^{i,j}\right|^2.
    \label{F_Norm}
\end{equation}
Thus $F_{N}(\xi)$ directly measures the total weight of the terms in $H_{eff}$ that are $\xi$-local in onsite $x$-bit-flip operators (in that sense, the above series is also ``locality-ordered" in $x$-basis, $\xi$ being the measure of the non-locality). If $F_{N}(\xi)$ falls exponentially (or faster) with $\xi$ beyond a scale that is independent of $L,$ then it will be called quasi-local. We now provide numerical evidence consistent with this picture. We compare the effective Hamiltonian structures for \eqref{eq:h2}, with and without the correction \eqref{eq:c_h2c}, in the strongly driven regime in Fig.~\ref{fig:contrast_LOE}.\\

\noindent
{\bf Conclusion:} We show that neither ergodicity nor its breaking is a universal consequence of non-integrability or strong drive in generic, disorder-free Floquet systems by analyzing the locality structure of the exact Floquet Hamiltonian. We show how local engineering can stabilize dynamical freezing and EB in systems where freezing is originally unstable. 
This broadens the scope for preserving useful quantum states and for DF-assisted quantum-error mitigation already demonstrated experimentally.
\\

{\bf Acknowledgment:}
We thank V. Khemani, R. Moessner, and D. Sen for related collaborations. We also thank V. Khemani especially for a careful reading of the manuscript and for providing many useful comments. AI assistance was used in editing and shortening the MS. No plotted simulation data were generated or altered by this assistance. AH acknowledges support from the European Union’s Horizon Europe Marie Sk\l{}odowska-Curie Action, under grant agreement No. 101110987-project CondmatQtech. 


\bibliography{Engineering_DF_V0}

\clearpage
\onecolumngrid
\clearpage

\setcounter{section}{0}
\setcounter{subsection}{0}
\setcounter{subsubsection}{0}
\setcounter{equation}{0}
\setcounter{figure}{0}
\setcounter{table}{0}
\renewcommand{\theequation}{S\arabic{equation}}
\renewcommand{\thefigure}{S\arabic{figure}}
\renewcommand{\thetable}{S\arabic{table}}
\renewcommand{\thesection}{S\Roman{section}}
\renewcommand{\thesubsection}{\thesection.\Alph{subsection}}
\renewcommand{\thesubsubsection}{\thesubsection.\arabic{subsubsection}}
\renewcommand{\theHequation}{SM.\arabic{equation}}
\renewcommand{\theHfigure}{SM.\arabic{figure}}
\renewcommand{\theHtable}{SM.\arabic{table}}
\renewcommand{\theHsection}{SM.\arabic{section}}
\renewcommand{\theHsubsection}{SM.\arabic{section}.\arabic{subsection}}

\begin{center}
{\large\bfseries Supplemental Material for:
Engineering Ergodicity-Breaking and Stable Emergent Conservation Laws in a Generic Quantum Many-body System\\
Under Strong Periodic Drive\par}
\end{center}
\vspace{1em}

In this Supplemental Material, we provide additional analytical derivations, numerical results, and methodological information supporting the results presented in the main text.

\section{DERIVATION OF EQ.~(7) OF THE MAIN TEXT}

We consider the general driven Hamiltonian introduced in the main text,
\begin{equation}
H(t)=H_{stat}+H_D^x(t),
\qquad
H_{stat}=H^x+V,
\qquad
H_D^x(t)=-A_D s(t)D^x,
\label{eq:SM_rot_start}
\end{equation}
where $D^x$ is the drive operator, $A_D$ is the
drive amplitude, and $s(t)$ is the periodic drive protocol. Defining
\begin{equation}
\chi(t)=\int_0^t s(u)\,du,
\end{equation}
the rotating-frame transformation is
\begin{equation}
R(t)=e^{iA_D\chi(t)D^x}.
\label{eq:SM_Rdef}
\end{equation}
The rotating-frame Hamiltonian is
\begin{align}
H^{\mathrm{Rot}}(t)
&=
R^\dagger(t)H(t)R(t)-iR^\dagger(t)\dot R(t)
\nonumber\\
&=
R^\dagger(t)H_{stat}R(t),
\label{eq:SM_Hrot_basic}
\end{align}
since
\begin{equation}
-iR^\dagger(t)\dot R(t)
=
A_D s(t)D^x
\end{equation}
cancels the drive term.

Let the spectral decomposition of $D^x$ be
\begin{equation}
D^x=\sum_\gamma \gamma P_\gamma,
\qquad
\sum_\gamma P_\gamma=\mathbb{I},
\qquad
P_\gamma P_{\gamma'}=\delta_{\gamma\gamma'}P_\gamma ,
\label{eq:SM_Dspectral}
\end{equation}
where $\gamma$ is an eigenvalue of $D^x$ and $P_\gamma$ projects onto
the corresponding eigenspace. Inserting the resolution of identity
on both sides of $H_{stat}$ gives
\begin{equation}
H_{stat}
=
\sum_{\gamma,\gamma'}
P_\gamma H_{stat}P_{\gamma'}.
\end{equation}
We group these matrix elements according to the change in drive
eigenvalue,
\begin{equation}
q=\gamma-\gamma',
\end{equation}
and define
\begin{equation}
V_q
=
\sum_{\substack{\gamma,\gamma'\\\gamma-\gamma'=q}}
P_\gamma H_{stat}P_{\gamma'}.
\label{eq:SM_Vqdef}
\end{equation}
This is Eq.~(6) of the main text. It follows directly that
\begin{align}
[D^x,V_q]
&=
\sum_{\substack{\gamma,\gamma'\\\gamma-\gamma'=q}}
(\gamma-\gamma')P_\gamma H_{stat}P_{\gamma'}
\nonumber\\
&=
qV_q.
\label{eq:SM_DVq}
\end{align}
In particular,
\begin{equation}
V_0=
\sum_\gamma P_\gamma H_{stat}P_\gamma.
\label{eq:SM_V0def}
\end{equation}
Since $[H^x,D^x]=0$, $V_0$ can equivalently be written as
\begin{equation}
V_0
=
H^x+\sum_\gamma P_\gamma V P_\gamma .
\label{eq:SM_V0explicit}
\end{equation}

The successive nested commutators are therefore
\begin{align}
[D^x,V_q]&=qV_q,\\
[D^x,[D^x,V_q]]&=q^2V_q,\\
[D^x,[D^x,[D^x,V_q]]]&=q^3V_q,\\
[D^x,[D^x,[D^x,[D^x,V_q]]]]&=q^4V_q.
\end{align}
Using the Baker--Campbell--Hausdorff expansion
\begin{equation}
e^{-X}Ye^{X}
=
Y-[X,Y]
+\frac{1}{2!}[X,[X,Y]]
-\frac{1}{3!}[X,[X,[X,Y]]]
+\frac{1}{4!}[X,[X,[X,[X,Y]]]]
+\cdots ,
\end{equation}
with
\begin{equation}
X=iA_D\chi(t)D^x,
\qquad
Y=V_q,
\end{equation}
we obtain
\begin{align}
R^\dagger V_qR
={}&
V_q
-iA_D\chi(t)[D^x,V_q]
-\frac{[A_D\chi(t)]^2}{2!}[D^x,[D^x,V_q]]
\nonumber\\
&+
\frac{i[A_D\chi(t)]^3}{3!}
[D^x,[D^x,[D^x,V_q]]]
\nonumber\\
&+
\frac{[A_D\chi(t)]^4}{4!}
[D^x,[D^x,[D^x,[D^x,V_q]]]]
+\cdots
\nonumber\\
={}&
\left[
1-iqA_D\chi(t)
-\frac{[qA_D\chi(t)]^2}{2!}
+\frac{i[qA_D\chi(t)]^3}{3!}
+\frac{[qA_D\chi(t)]^4}{4!}
+\cdots
\right]V_q
\nonumber\\
={}&
e^{-iqA_D\chi(t)}V_q.
\label{eq:SM_Vq_rotation}
\end{align}
Since
\begin{equation}
H_{stat}=\sum_qV_q,
\end{equation}
Eq.~(\ref{eq:SM_Hrot_basic}) becomes
\begin{align}
H^{\mathrm{Rot}}(t)
&=
\sum_q e^{-iqA_D\chi(t)}V_q
\nonumber\\
&=
V_0+
\sum_{q\neq0}
e^{-iqA_D\chi(t)}V_q.
\label{eq:SM_rot_final}
\end{align}
This is Eq.~(7) of the main text.

\section{STABILITY CRITERION AND THE SPECIAL ROLE OF $V_0$}

Equation~(7) of the main text gives the rotating-frame Hamiltonian as
\begin{equation}
H^{\mathrm{Rot}}(t)
=
V_0+\sum_{q\neq0}
e^{-iqA_D\chi(t)}V_q ,
\label{eq:SM_Hrot_decomp}
\end{equation}
with
\begin{equation}
V_0
=
\sum_\gamma P_\gamma H_{stat}P_\gamma
=
H^x+\sum_\gamma P_\gamma V P_\gamma ,
\label{eq:SM_V0}
\end{equation}
where $P_\gamma$ projects onto the eigenspace of the drive operator
$D^x$ with eigenvalue $\gamma$.

The special role of $V_0$ follows directly from the fact that it
commutes with the drive operator. Using
\begin{equation}
D^xP_\gamma=P_\gamma D^x=\gamma P_\gamma ,
\end{equation}
we obtain
\begin{align}
[D^x,V_0]
&=
\sum_\gamma
\left(
D^xP_\gamma H_{stat}P_\gamma
-
P_\gamma H_{stat}P_\gamma D^x
\right)
\nonumber\\
&=
\sum_\gamma
\left(
\gamma P_\gamma H_{stat}P_\gamma
-
\gamma P_\gamma H_{stat}P_\gamma
\right)
=0 .
\label{eq:SM_DV0}
\end{align}
Thus every contribution contained in $V_0$ remains explicitly time independent in the rotating frame and survives with its bare $O(1)$ strength, without suppression by the large drive amplitude $A_D$. 

Given the stability criterion stated in the main text, namely that $V_0$ must be composed exclusively of $X$ operators, any contributions involving $Y$ or $Z$ operators in $V_0$ must be removed explicitly, since they survive with their bare $O(1)$ strength and are not suppressed by the strong drive. By contrast, the contributions to $H_{eff}$ from terms involving
$Y$ or $Z$ operators in the $q\neq0$ sectors are suppressed by inverse powers of the large drive amplitude $A_D$.

Accordingly, stable DF can be engineered either by modifying the time-independent Hamiltonian to remove the unwanted projected contribution,
\begin{equation}
\delta H_{stat}
=
-\sum_\gamma P_\gamma V P_\gamma ,
\label{eq:SM_static_engineering}
\end{equation}
so that $V_0^{\rm eng}=H^x$, or by modifying the drive operator so that the unwanted quantum terms lie entirely in the $q\neq0$ sectors.

\section{DERIVATION OF THE $V_0$ SECTORS AND ENGINEERING PRESCRIPTIONS}

Here we derive the $q=0$ sectors $V_0$ for the models introduced in
the main text and show explicitly how the two engineering prescriptions
eliminate the unwanted $Y$- and $Z$-containing contributions from
$V_0$.

\subsection{Uniform field drive: $H_1$}

For
\begin{equation}
H_1(t)
=
H_1^x-h^z\sum_i Z_i
-h_D^x s_1(t)\sum_iX_i ,
\end{equation}
the drive operator is
\begin{equation}
D^x=\sum_iX_i .
\end{equation}
Since $H_1^x$ commutes with $D^x$, it belongs entirely to $V_0$.

To determine the projection of the transverse-field term, we introduce
the local projectors
\begin{equation}
P_i^\pm=\frac{1\pm X_i}{2}.
\label{eq:SM_H1_projectors}
\end{equation}
Since $Z_i$ flips the $x$ spin at site $i$, it changes the eigenvalue
of $D^x$ by $\pm2$, and hence
\begin{equation}
P_i^\pm Z_iP_i^\pm=0.
\end{equation}
The transverse field therefore has no $q=0$ component, giving
\begin{equation}
V_{0,1}=H_1^x .
\label{eq:SM_H1V0}
\end{equation}
Thus $V_{0,1}$ is composed exclusively of $X$ operators.

\subsection{Uniform bond drive: $H_2$}

For
\begin{equation}
H_2(t)
=
H_2^x-h^z\sum_iZ_i
-J_Ds_1(t)\sum_iX_iX_{i+1},
\end{equation}
the drive operator is
\begin{equation}
D^x=\sum_iX_iX_{i+1}.
\end{equation}
As in the main text, the $q=0$ projection of the transverse-field term
can be resolved locally using
\begin{equation}
P_i^\pm
=
\frac{1\pm X_{i-1}X_{i+1}}{2}.
\label{eq:SM_H2_projectors}
\end{equation}
A flip of spin $i$ reverses the two adjacent bond variables
$X_{i-1}X_i$ and $X_iX_{i+1}$. The drive eigenvalue remains unchanged
only when the neighboring spins are oppositely aligned. Therefore,
\begin{equation}
\sum_\gamma P_\gamma Z_iP_\gamma
=
Z_iP_i^-
=
\frac{1}{2}
\left(
Z_i-X_{i-1}Z_iX_{i+1}
\right),
\label{eq:SM_H2_Z_q0}
\end{equation}
and hence
\begin{align}
V_{0,2}
&=
H_2^x-h^z\sum_iZ_iP_i^-
\nonumber\\
&=
H_2^x
-\frac{h^z}{2}\sum_i
\left(
Z_i-X_{i-1}Z_iX_{i+1}
\right).
\label{eq:SM_H2V0}
\end{align}
Thus $V_{0,2}$ contains $Z$ operators and does not satisfy the
stability criterion.

To eliminate this contribution from $V_{0,2}$, we add the
time-independent correction
\begin{align}
H_2^C
&=
h^z\sum_iZ_iP_i^-
\nonumber\\
&=
\frac{h^z}{2}\sum_i
\left(
Z_i-X_{i-1}Z_iX_{i+1}
\right),
\label{eq:SM_H2C}
\end{align}
which gives
\begin{equation}
V_{0,2}^{\rm eng}=H_2^x .
\end{equation}
The engineered $V_{0,2}$ is therefore composed exclusively of $X$
operators.

\subsection{Uniform field drive with $ZZ$ interactions: $H_3$}

For
\begin{equation}
H_3(t)
=
H_3^x
-\sum_{r=1}^{2}J_r^z\sum_i Z_iZ_{i+r}
-h_D^xs_2(t)\sum_iX_i ,
\end{equation}
with
\begin{equation}
H_3^x=g^x\sum_iX_i,
\end{equation}
the drive operator is again
\begin{equation}
D^x=\sum_iX_i .
\end{equation}

A term $Z_iZ_{i+r}$ flips the two $x$ spins at sites $i$ and $i+r$.
Introducing
\begin{equation}
P_i^\pm=\frac{1\pm X_i}{2},
\qquad
P_{i+r}^\pm=\frac{1\pm X_{i+r}}{2},
\label{eq:SM_H3_projectors}
\end{equation}
the $q=0$ sector corresponds to initially anti-aligned spins, for
which the simultaneous flip leaves the eigenvalue of $D^x$ unchanged.
Thus
\begin{align}
\sum_\gamma
P_\gamma Z_iZ_{i+r}P_\gamma
={}&
P_i^+P_{i+r}^-Z_iZ_{i+r}P_i^-P_{i+r}^+
\nonumber\\
&+
P_i^-P_{i+r}^+Z_iZ_{i+r}P_i^+P_{i+r}^-
\nonumber\\
={}&
\frac{1}{2}
\left(
Z_iZ_{i+r}+Y_iY_{i+r}
\right).
\end{align}
Consequently,
\begin{equation}
V_{0,3}
=
H_3^x
-\frac{1}{2}\sum_{r=1}^{2}J_r^z\sum_i
\left(
Z_iZ_{i+r}+Y_iY_{i+r}
\right).
\label{eq:SM_H3V0}
\end{equation}
Thus $V_{0,3}$ contains both $Y$ and $Z$ operators and does not satisfy
the stability criterion.

The required time-independent correction is
\begin{equation}
H_3^C
=
+\sum_{r=1}^{2}J_r^z\sum_iY_iY_{i+r},
\label{eq:SM_H3C}
\end{equation}
or, equivalently,
\begin{equation}
J_1^y=-J_1^z,
\qquad
J_2^y=-J_2^z .
\end{equation}
The resulting interaction is
\begin{equation}
V_3^{\rm mod}
=
-\sum_{r=1}^{2}J_r^z\sum_i
\left(
Z_iZ_{i+r}-Y_iY_{i+r}
\right),
\label{eq:SM_H3_modified}
\end{equation}
whose $q=0$ projection vanishes. Therefore,
\begin{equation}
V_{0,3}^{\rm eng}
=
H_3^x
=
g^x\sum_iX_i .
\end{equation}
The engineered $V_{0,3}$ is thus composed exclusively of $X$
operators.

\subsection{Engineered bond drive: $H_2$ with $D_2^x$}

We next keep the time-independent Hamiltonian unchanged and instead
modify the drive operator. For $H_2$, the engineered drive is
\begin{equation}
D_2^x
=
\sum_i
\left[
J_D+(-1)^{i-1}J_{SD}
\right]
X_iX_{i+1}.
\label{eq:SM_D2x}
\end{equation}
A spin flip at site $i$ reverses the two adjacent bond variables.
If the two bonds have the same eigenvalue, the change in the drive
eigenvalue is $\pm4J_D$, whereas for opposite bond eigenvalues it is
$\pm4J_{SD}$. Therefore, when both $J_D$ and $J_{SD}$ are nonzero,
the transverse-field term has no zero drive-eigenvalue component.
Hence,
\begin{equation}
V_{0,2}^{D_2}=H_2^x .
\label{eq:SM_H2D2_V0}
\end{equation}
Thus the engineered drive transfers the entire transverse-field
contribution out of the $q=0$ sector, leaving $V_{0,2}^{D_2}$
composed exclusively of $X$ operators.

\subsection{Engineered field drive: $H_3$ with $D_3^x$}

For $H_3$, the engineered drive operator is
\begin{equation}
D_3^x
=
\sum_i a_iX_i,
\qquad
a_i
=
h_D^x
+h_{SD,1}^x(-1)^{i-1}
+h_{SD,2}^x\eta_i,
\label{eq:SM_D3x}
\end{equation}
where
\begin{equation}
\eta_i=(-1)^{\lfloor(i-1)/2\rfloor}.
\end{equation}

A term $Z_iZ_{i+r}$ flips the two $x$ spins at sites $i$ and $i+r$.
For initially aligned spins, the corresponding change in the drive
eigenvalue is
\begin{equation}
\Delta d=\pm2(a_i+a_{i+r}),
\end{equation}
whereas for initially anti-aligned spins it is
\begin{equation}
\Delta d=\pm2(a_i-a_{i+r}).
\end{equation}
Thus, if
\begin{equation}
a_i\neq\pm a_{i+r},
\qquad
r=1,2,
\label{eq:SM_D3_nonzero_condition}
\end{equation}
for every interacting pair, none of the $ZZ$ pair-flip processes has
a zero drive-eigenvalue change. Consequently,
\begin{equation}
V_{0,3}^{D_3}
=
H_3^x
=
g^x\sum_iX_i .
\label{eq:SM_H3D3_V0}
\end{equation}

For the choice used in the main text,
\begin{equation}
h_{SD,1}^x=\frac{3}{4}h_D^x,
\qquad
h_{SD,2}^x=\frac{1}{2}h_D^x,
\end{equation}
the local fields repeat with the four-site pattern
\begin{equation}
(a_1,a_2,a_3,a_4)
=
\left(
\frac94,\frac34,\frac54,-\frac14
\right)h_D^x .
\label{eq:SM_D3_pattern}
\end{equation}
For both $r=1$ and $r=2$, all combinations
$a_i\pm a_{i+r}$ are nonzero, and hence
Eq.~(\ref{eq:SM_D3_nonzero_condition}) is satisfied for every interacting pair. The engineered $V_{0,3}^{D_3}$ is therefore composed exclusively of $X$ operators.


\section{ADDITIONAL NUMERICAL DETAILS: PARAMETER VALUES AND MIXED-STATE PREPARATION}

\subsection{Symmetry sectors}

All numerical results presented in the main text and the Supplemental Material were obtained within appropriate symmetry sectors. For the demonstration of non-universality and for the perturbation-engineered versions of both $H_2$ and $H_3$, we work in the $k=0$, $p=+1$ sector, where $k$ denotes the lattice momentum and $p$ the reflection parity about the center of the chain.

For the drive-engineered Ising model, results have been obtained in the zero-momentum sector $k=0$, defined with respect to translations by two sites. For the drive-operator-engineered version of $H_3$, we similarly use the $k=0$ sector defined with respect to translations by four sites.

For the results presented in the subsection on the Spin-flip Ordered Series, no particular translation or reflection symmetry sector has been selected. Instead, the effective Hamiltonian and the corresponding quantities have been constructed directly in the full Hilbert space of dimension $2^L$, since the spin-flip order is defined directly between individual $x$-bit-string configurations.
\subsection{Preparation of thermal mixed state}

We define the design Hamiltonian
\begin{equation}
H_{\mathrm{Des}}
=
-J_{\mathrm{Des}}\sum_i X_iX_{i+1}
+\kappa_{\mathrm{Des}}\sum_i X_iX_{i+2}
-h_{\mathrm{Des}}^x\sum_i X_i
-h_{\mathrm{Des}}^z\sum_i Z_i ,
\end{equation}
with
$J_{\mathrm{Des}}=1.0$, $\kappa_{\mathrm{Des}}=0.7$,
$h_{\mathrm{Des}}^x=5.1$, and $h_{\mathrm{Des}}^z=1.2$.
The thermal density matrix is constructed from the eigenstates
$|\epsilon_j\rangle$ of $H_{\mathrm{Des}}$ as
\begin{equation}
\rho_{\mathrm{Th}}(0)
=
\sum_{j=1}^{D_H}
\frac{e^{-\beta\epsilon_j}}{\mathcal Z}
|\epsilon_j\rangle\langle\epsilon_j|,
\qquad
\mathcal Z=\sum_{j=1}^{D_H}e^{-\beta\epsilon_j},
\end{equation}
where $D_H$ is the Hilbert-space dimension of the symmetry sector under consideration. We use $\beta=0.05$.

Let $|\mu_k\rangle$ denote the Floquet eigenstates,
$U(T,0)|\mu_k\rangle=e^{-i\mu_k}|\mu_k\rangle$. The corresponding
diagonal-ensemble density matrix is
\begin{equation}
\rho_{\mathrm{DE}}
=
\sum_k
\left(
\sum_{j=1}^{D_H}
\frac{e^{-\beta\epsilon_j}}{\mathcal Z}
|\langle\epsilon_j|\mu_k\rangle|^2
\right)
|\mu_k\rangle\langle\mu_k|.
\end{equation}
The diagonal-ensemble average (DEA) value, corresponding to the
infinite-time stroboscopic average, of an operator $O$ is then
\begin{equation}
\langle O\rangle_{\mathrm{DEA}}
=
\operatorname{Tr}\!\left(\rho_{\mathrm{DE}}O\right).
\end{equation}


\section{ADDITIONAL NUMERICAL RESULTS}

\subsection{Non-universality: Real-time dynamics, amplitude scans, and system-size dependence}

\begin{figure}[!htbp]
\centering
\includegraphics[width=0.325\textwidth]
{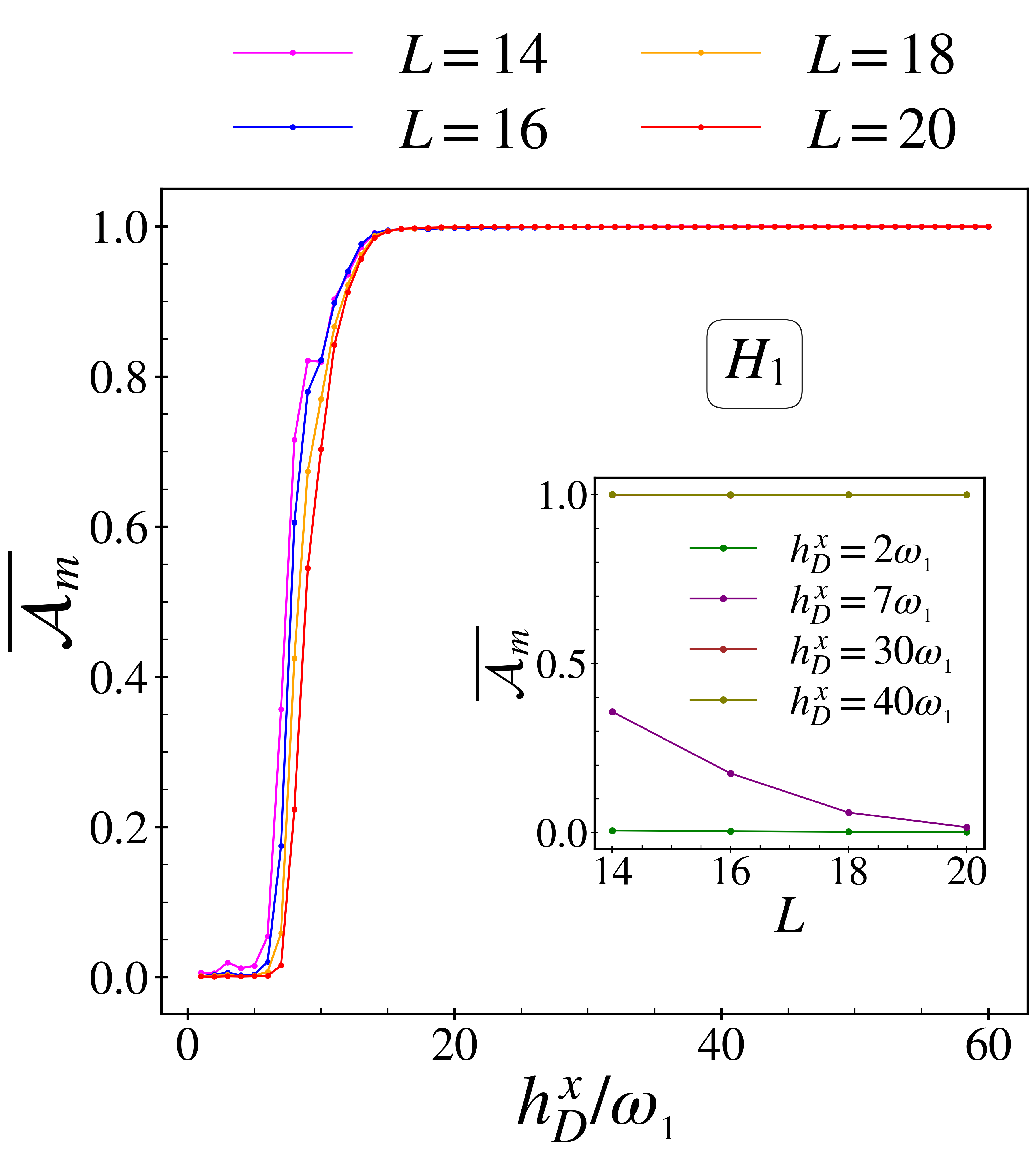}
\hfill
\includegraphics[width=0.325\textwidth]
{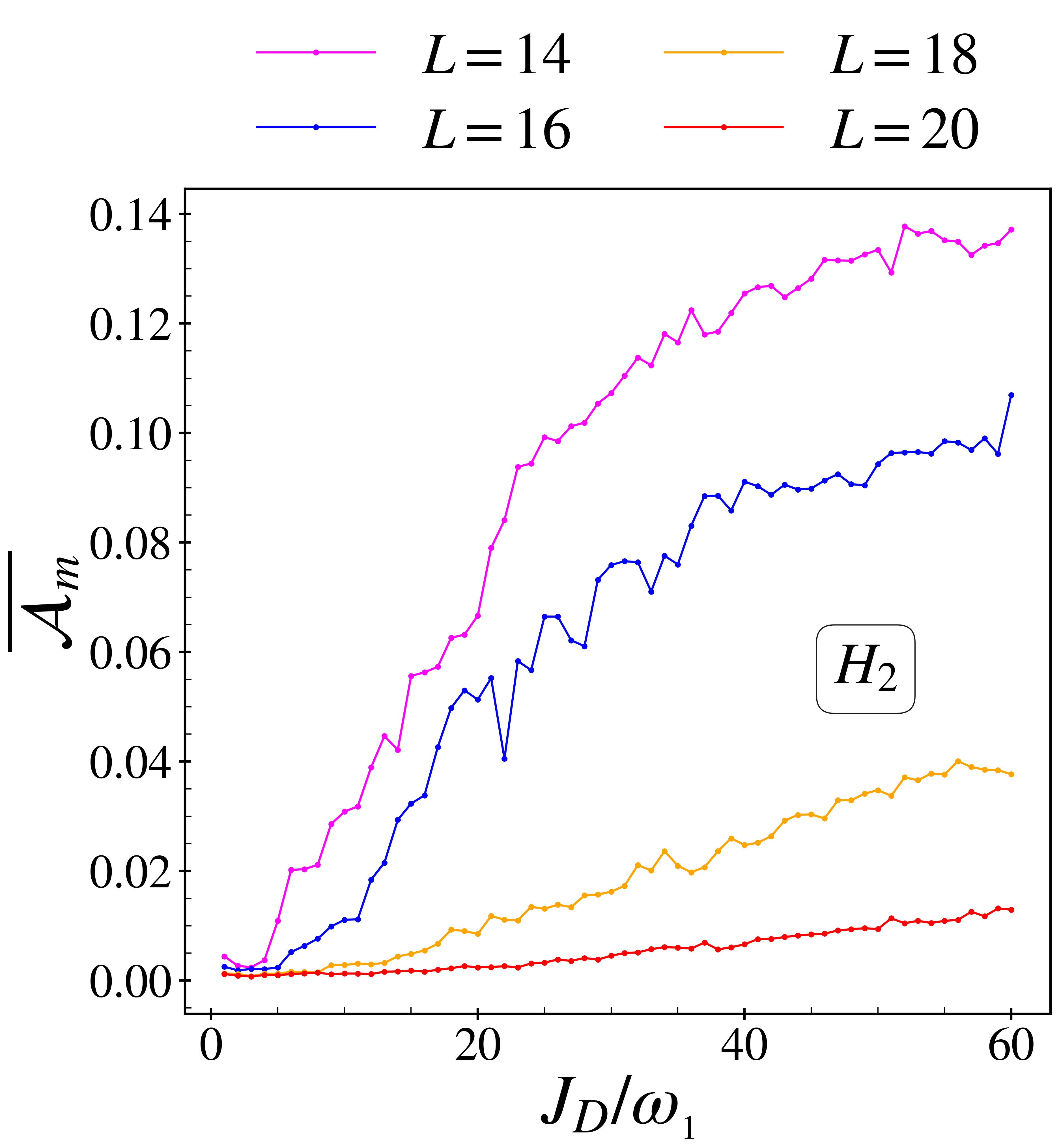}
\hfill
\includegraphics[width=0.325\textwidth]
{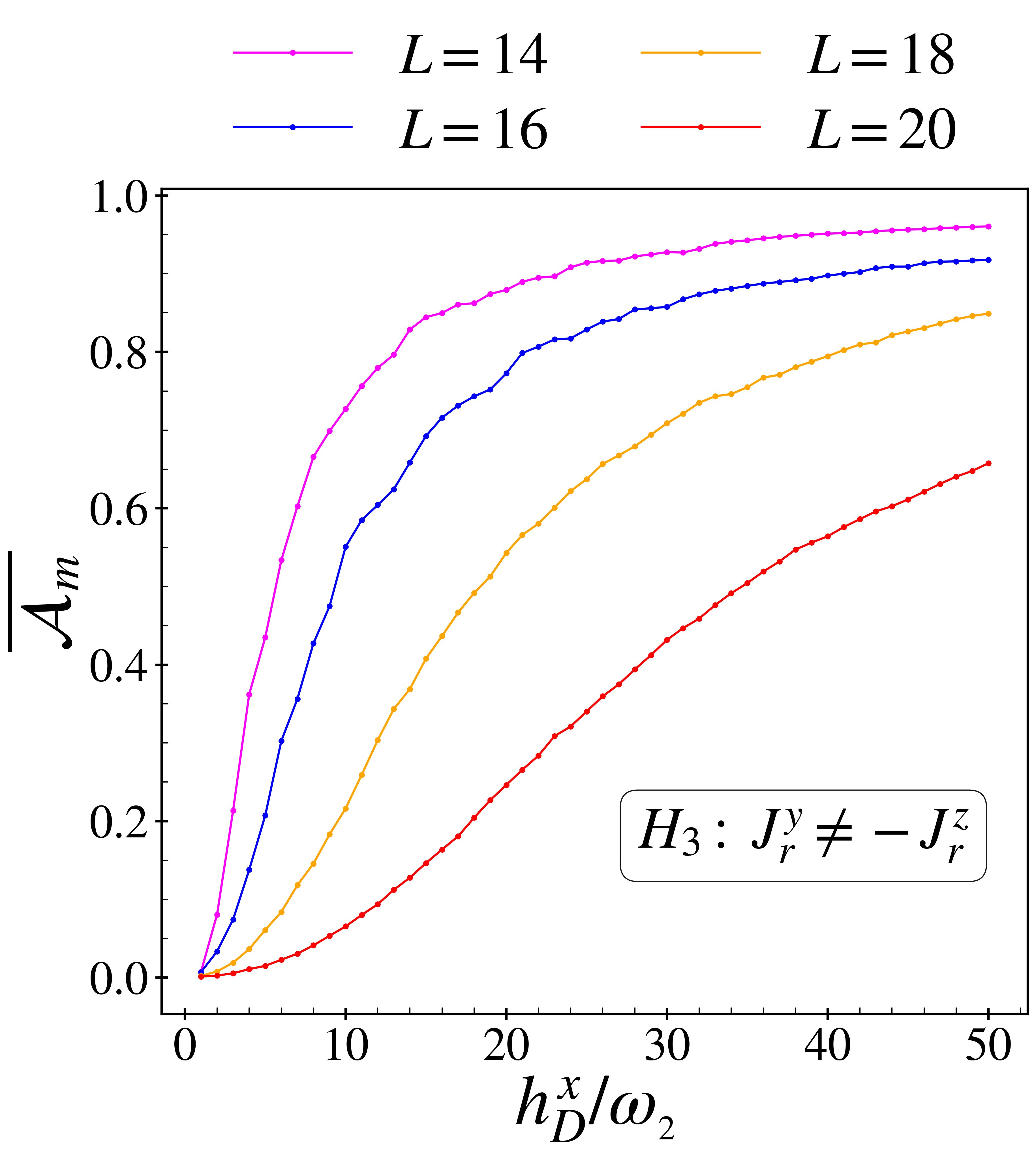}
\caption{
\textbf{Drive-amplitude dependence of the normalized
long-time-averaged autocorrelation of $m^x$ for different system
sizes.}
\textbf{(Left)} $H_1$: the systematic decrease with increasing $L$
at low drive disappears beyond a finite drive strength; the inset shows the $L$ dependence at selected drive amplitudes.
\textbf{(Center)} $H_2$ and
\textbf{(Right)} $H_3$: the systematic decrease with increasing $L$ persists throughout the examined ranges.
For all panels, the long-time average is computed over $1000$ uniformly spaced values of $n$ in $10^{8}\leq n\leq10^{11}$, where $n$ denotes the number of drive cycles.
}
\label{fig:SM_nonuniversality_amplitude_scan}
\end{figure}

Figure~\ref{fig:SM_nonuniversality_amplitude_scan} shows the system-size dependence of the normalized long-time-averaged autocorrelation of $m^x$ as a function of drive amplitude. At low drive amplitudes, it decreases systematically with increasing $L$
for all three Hamiltonians. For $H_1$, this systematic size dependence disappears beyond a finite drive strength and the autocorrelation remains close to unity. In contrast, for $H_2$ and $H_3$, the decreasing trend with increasing $L$ persists throughout the examined drive ranges.

\begin{figure}[!htbp]
\centering
\includegraphics[width=0.325\textwidth]
{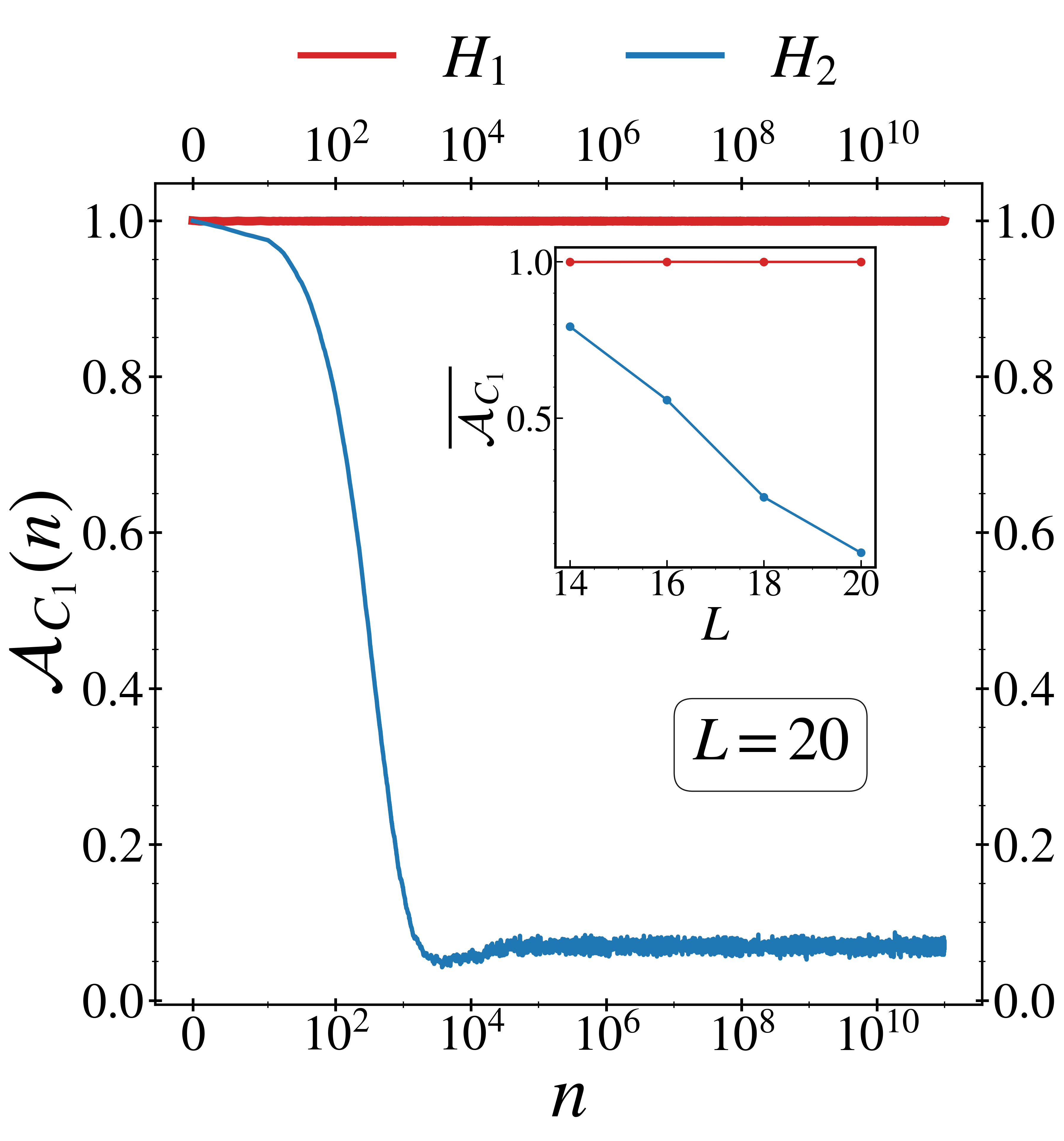}
\hfill
\includegraphics[width=0.325\textwidth]
{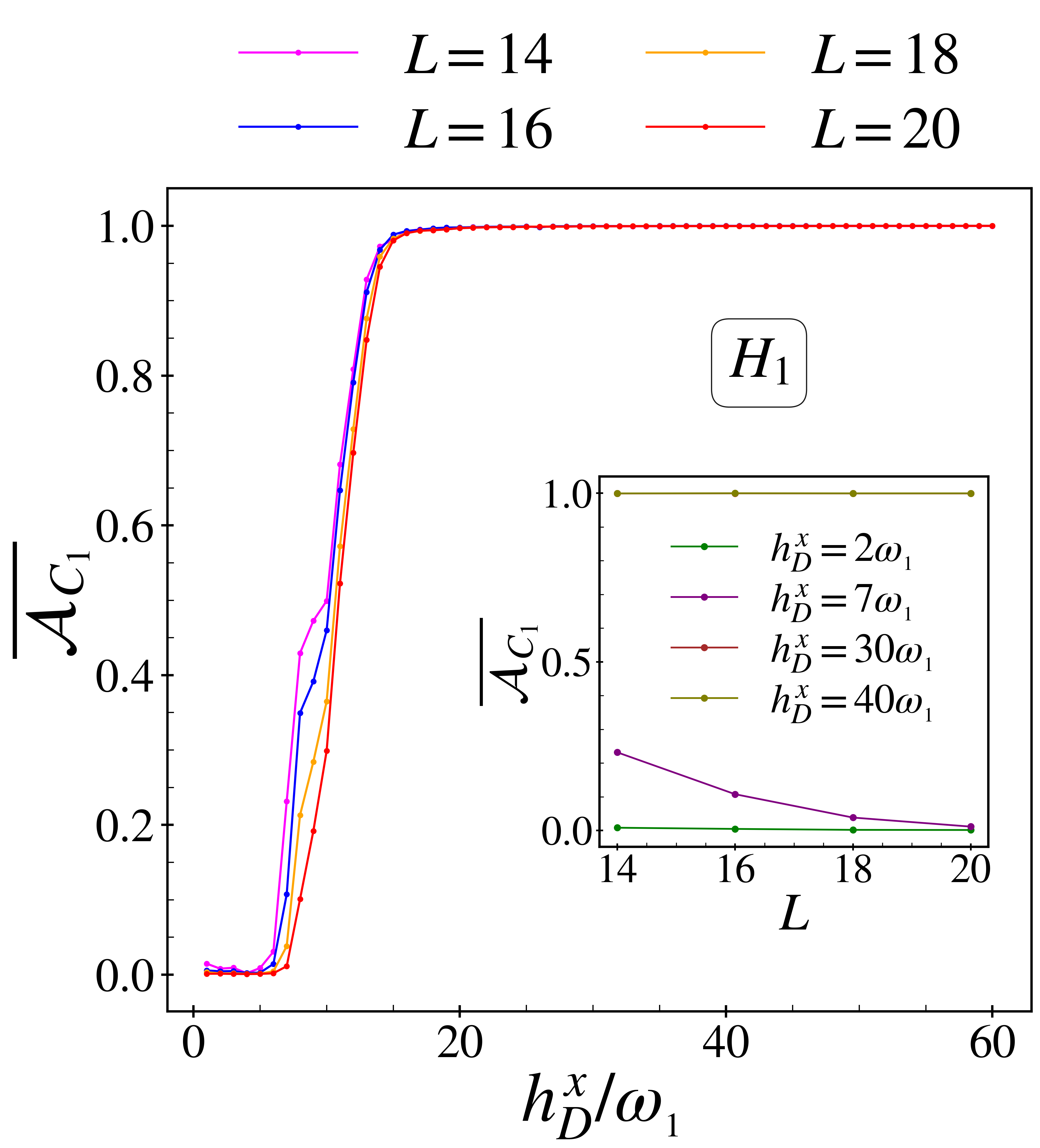}
\hfill
\includegraphics[width=0.325\textwidth]
{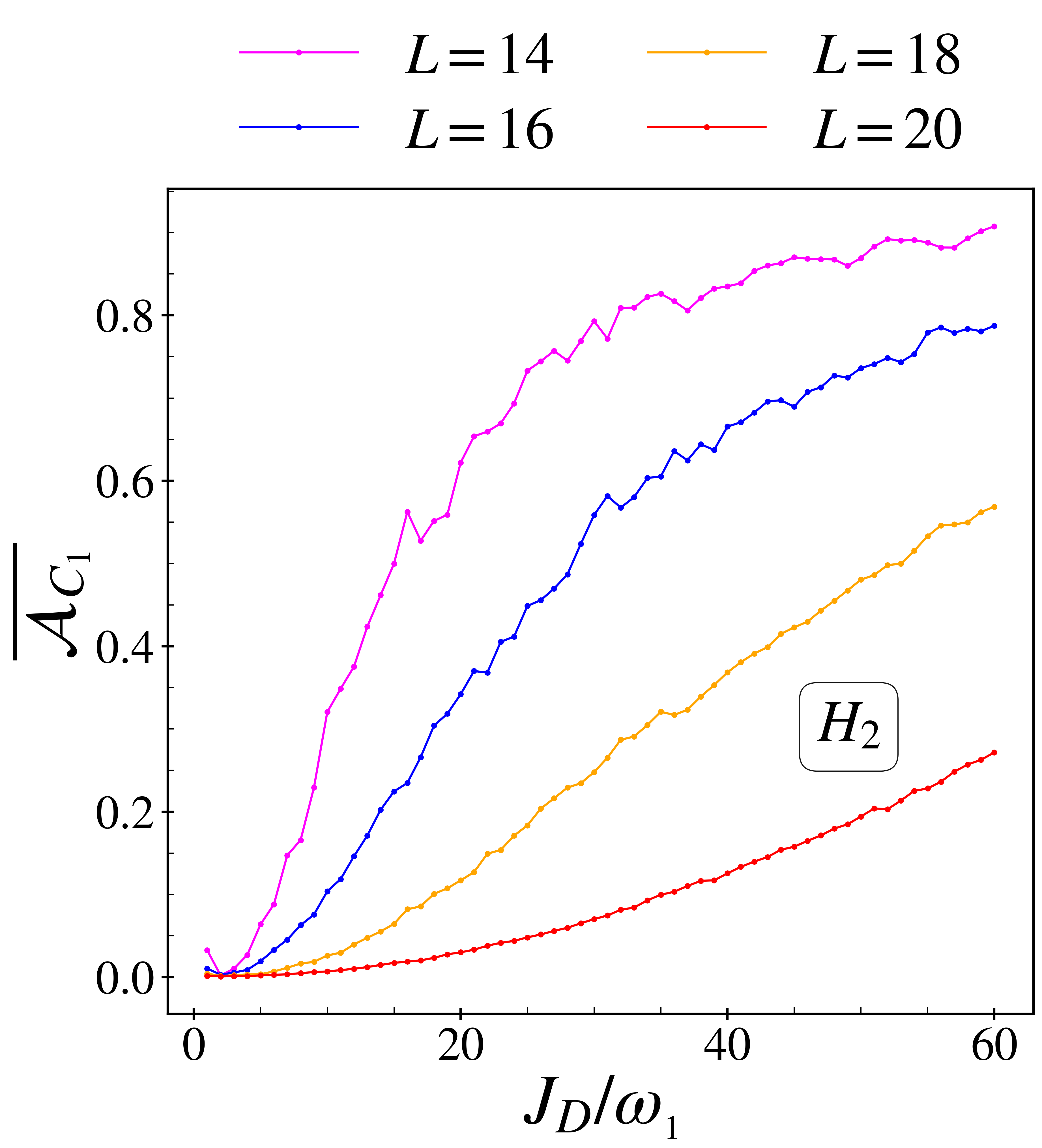}
\caption{
\textbf{Autocorrelation of the nearest-neighbor correlator $C_1^x$
for $H_1$ and $H_2$.}
\textbf{(Left)} Strong-drive dynamics: the autocorrelation remains
close to unity for $H_1$ but decays substantially for $H_2$.
\textbf{(Center)} Drive-amplitude scan for $H_1$: the systematic
decrease with increasing $L$ disappears at sufficiently strong drive.
\textbf{(Right)} Corresponding scan for $H_2$, where the decrease
with increasing $L$ persists throughout the examined range.
}
\label{fig:SM_nonuniversality_C1x}
\end{figure}

The same contrast extends to the short-range correlators
\begin{equation}
C_r^x=L^{-1}\sum_iX_iX_{i+r},
\qquad r=1,2.
\end{equation}
For $C_1^x$, Fig.~\ref{fig:SM_nonuniversality_C1x} shows that its
strong-drive autocorrelation remains close to unity for $H_1$ but
decays substantially for $H_2$. Correspondingly, the systematic
decrease of its long-time average with increasing $L$ disappears at
strong drive for $H_1$ but persists for $H_2$.

\begin{figure}[!htbp]
\centering
\includegraphics[width=0.325\textwidth]
{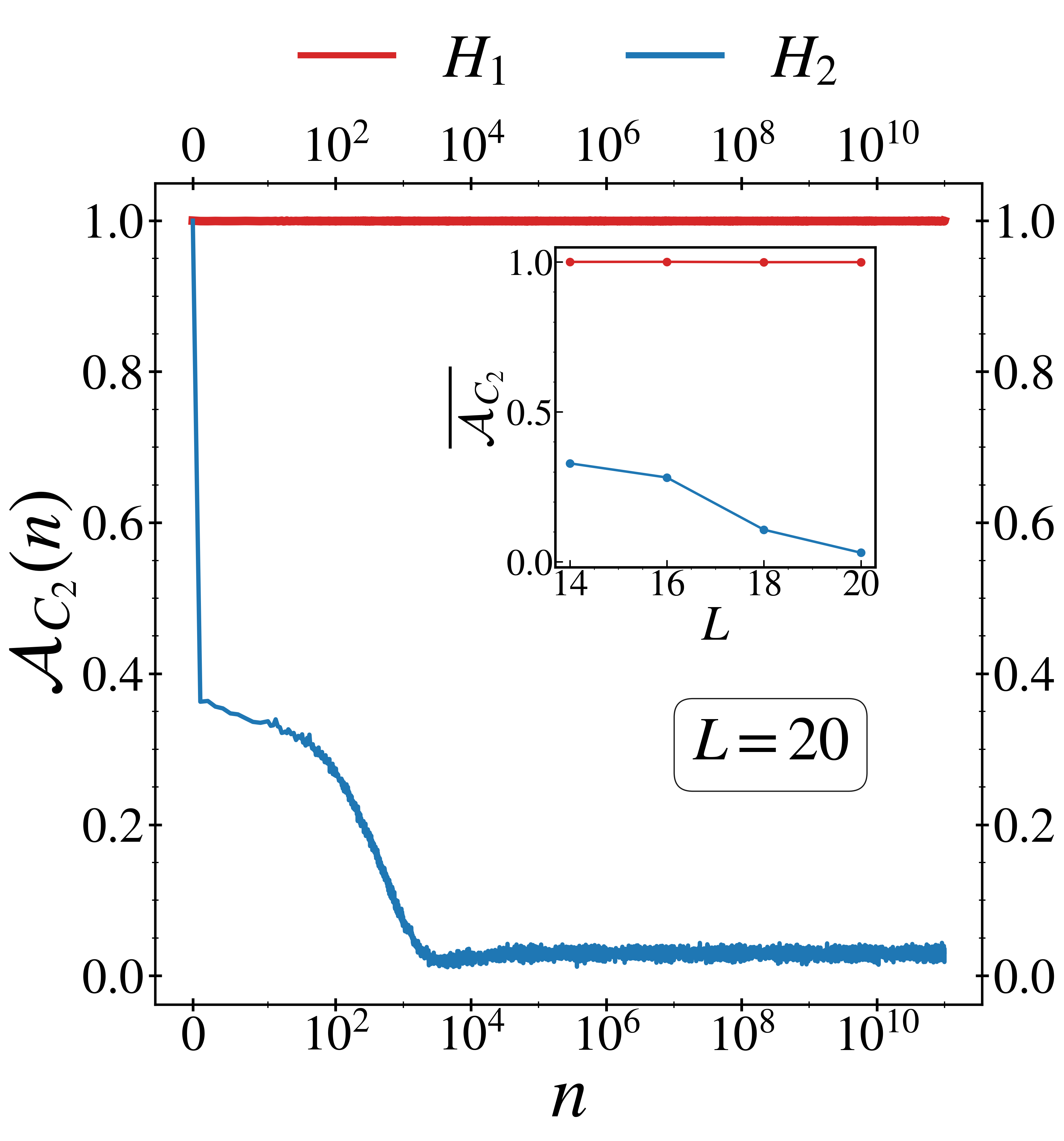}
\hfill
\includegraphics[width=0.325\textwidth]
{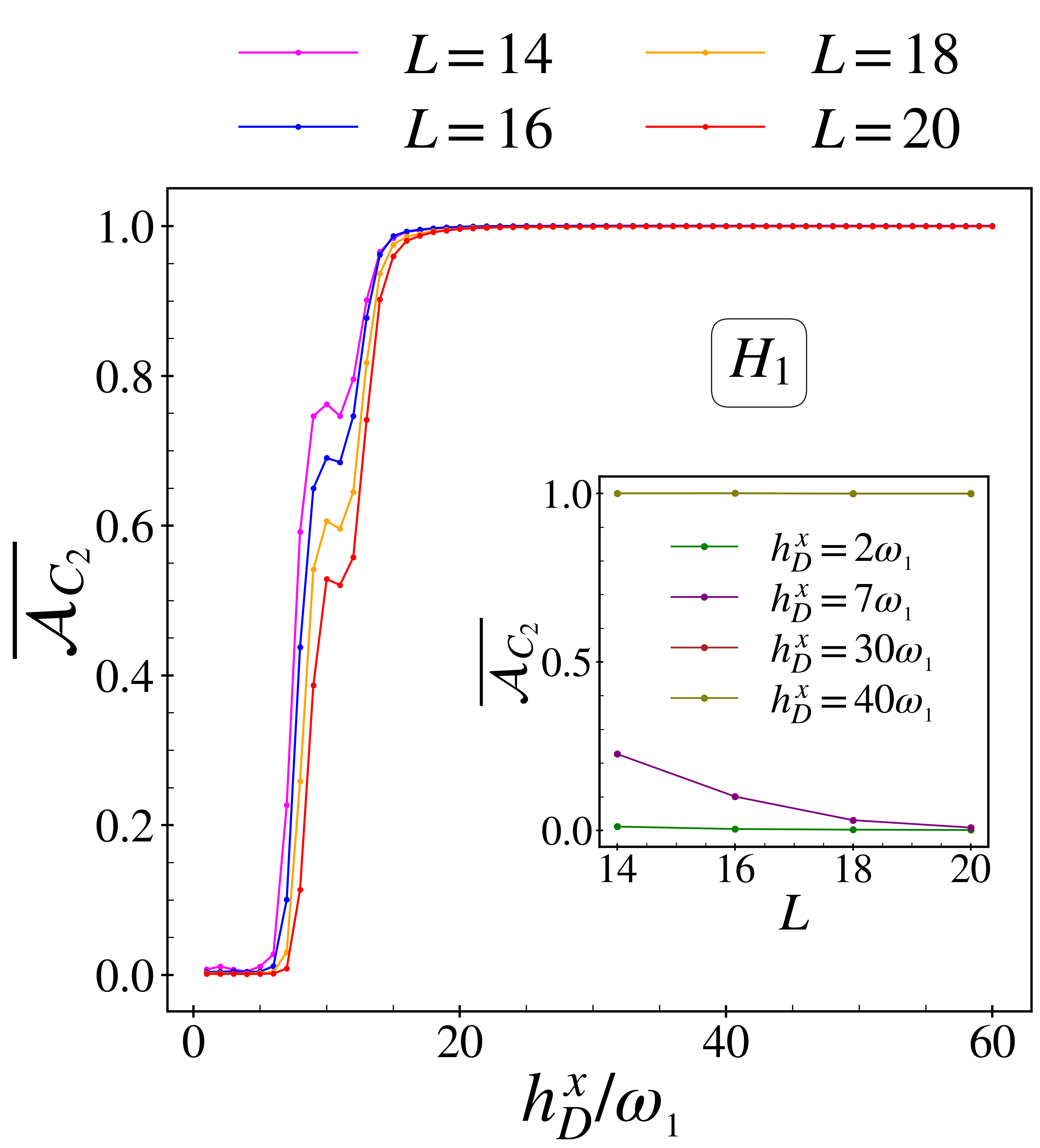}
\hfill
\includegraphics[width=0.325\textwidth]
{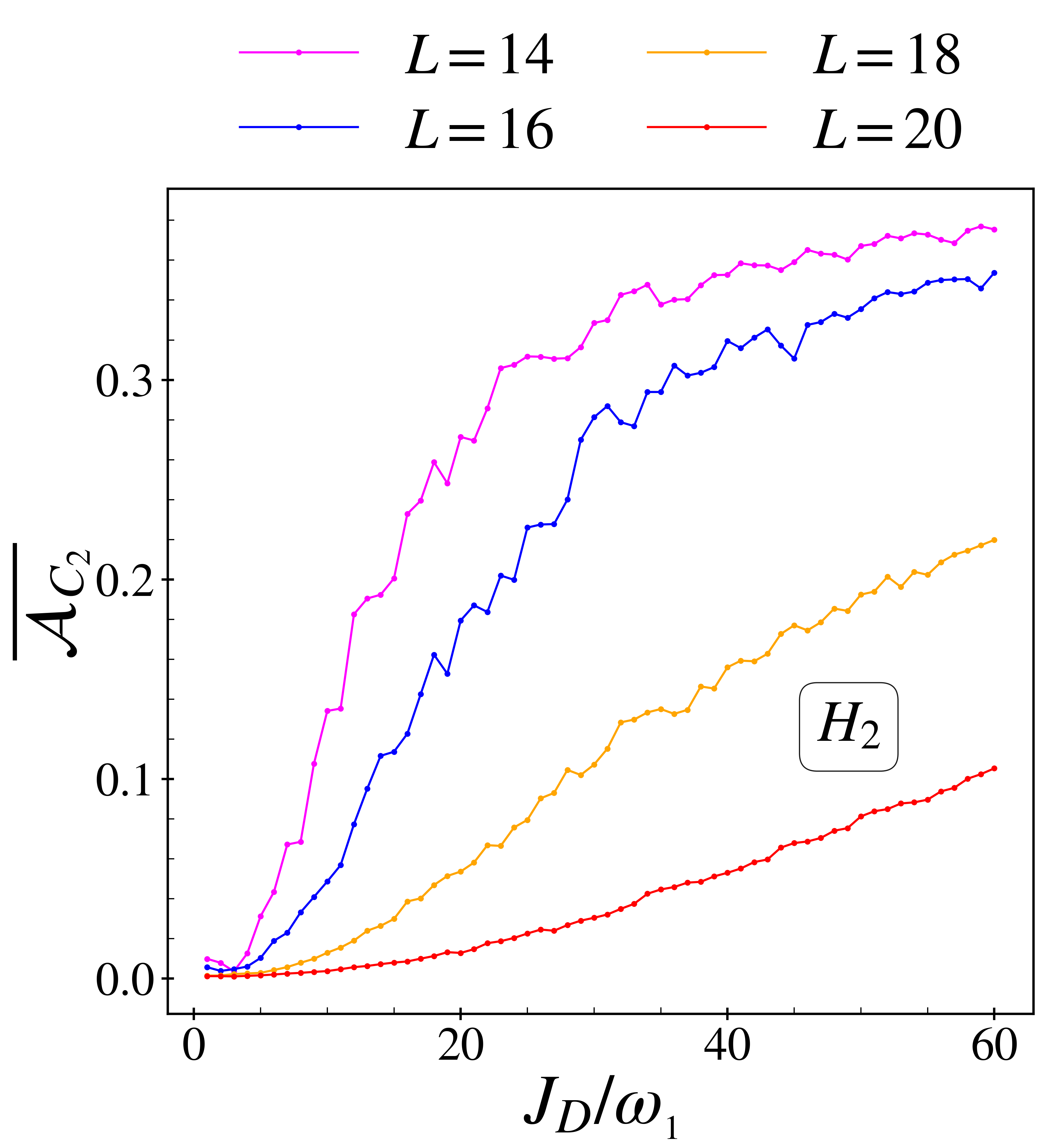}
\caption{
\textbf{Autocorrelation of the next-nearest-neighbor correlator
$C_2^x$ for $H_1$ and $H_2$.}
\textbf{(Left)} Strong-drive dynamics: the autocorrelation remains
close to unity for $H_1$ but decays for $H_2$.
\textbf{(Center)} Drive-amplitude scan for $H_1$: the systematic
decrease with increasing $L$ disappears at sufficiently strong drive.
\textbf{(Right)} Corresponding scan for $H_2$, where the decrease
with increasing $L$ persists throughout the examined range.
}
\label{fig:SM_nonuniversality_C2x}
\end{figure}

Figure~\ref{fig:SM_nonuniversality_C2x} shows the same behavior for
$C_2^x$: its strong-drive autocorrelation remains close to unity for
$H_1$ but decays for $H_2$, while the systematic size dependence of
the long-time average disappears only for $H_1$.

\subsection{Time-independent stabilization by Hamiltonian engineering:
Real-time dynamics, amplitude scans, and system-size dependence}

We first present additional numerical results for the
time-independent stabilization route, in which the time-independent
part of the laboratory-frame Hamiltonian is engineered so that $V_0$
becomes classical.

\begin{figure}[!htbp]
\centering
\includegraphics[width=0.325\textwidth]
{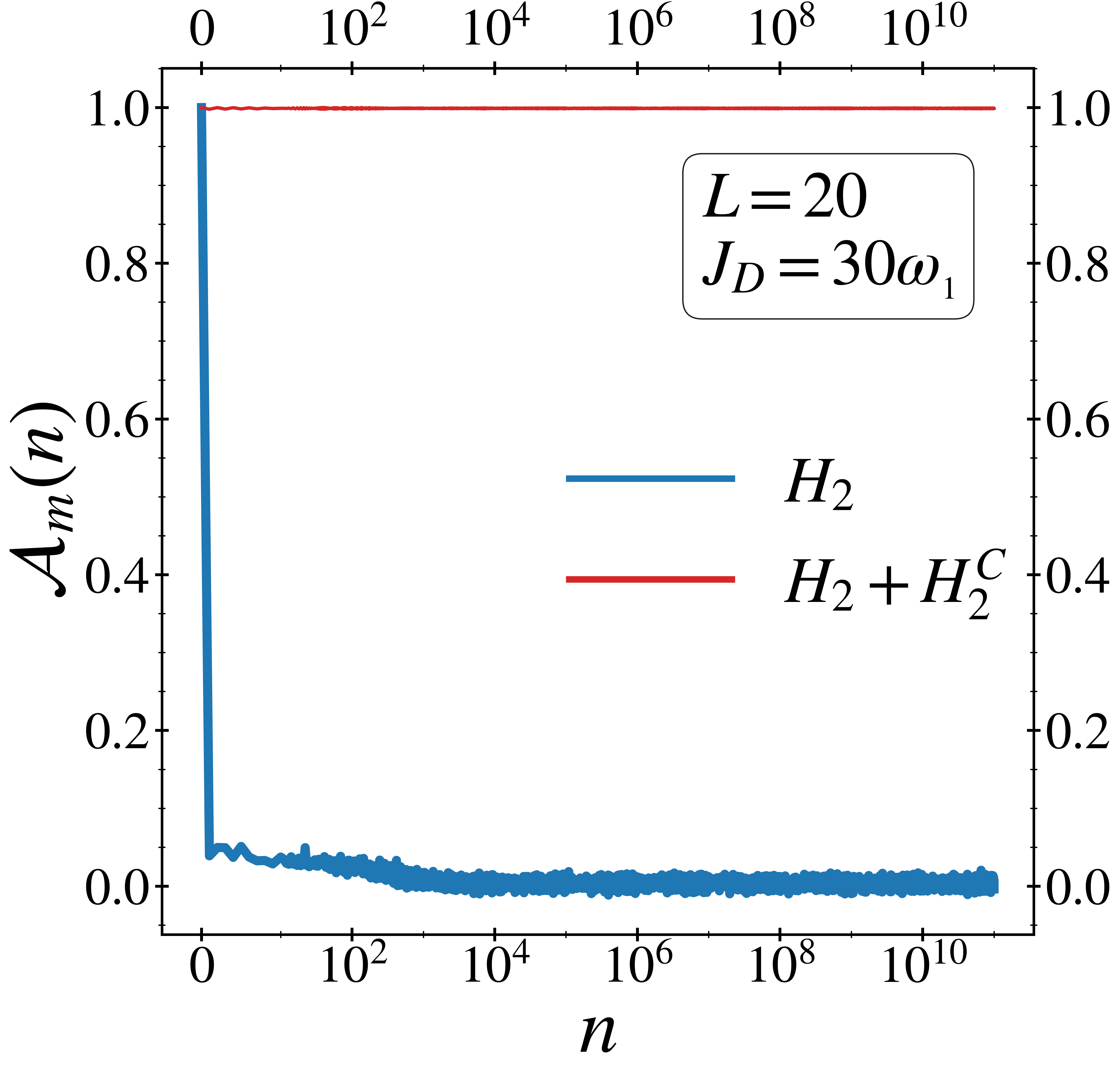}
\hfill
\includegraphics[width=0.325\textwidth]
{Fig_DriveScan_H2_Atcr_mx_2.png}
\hfill
\includegraphics[width=0.325\textwidth]
{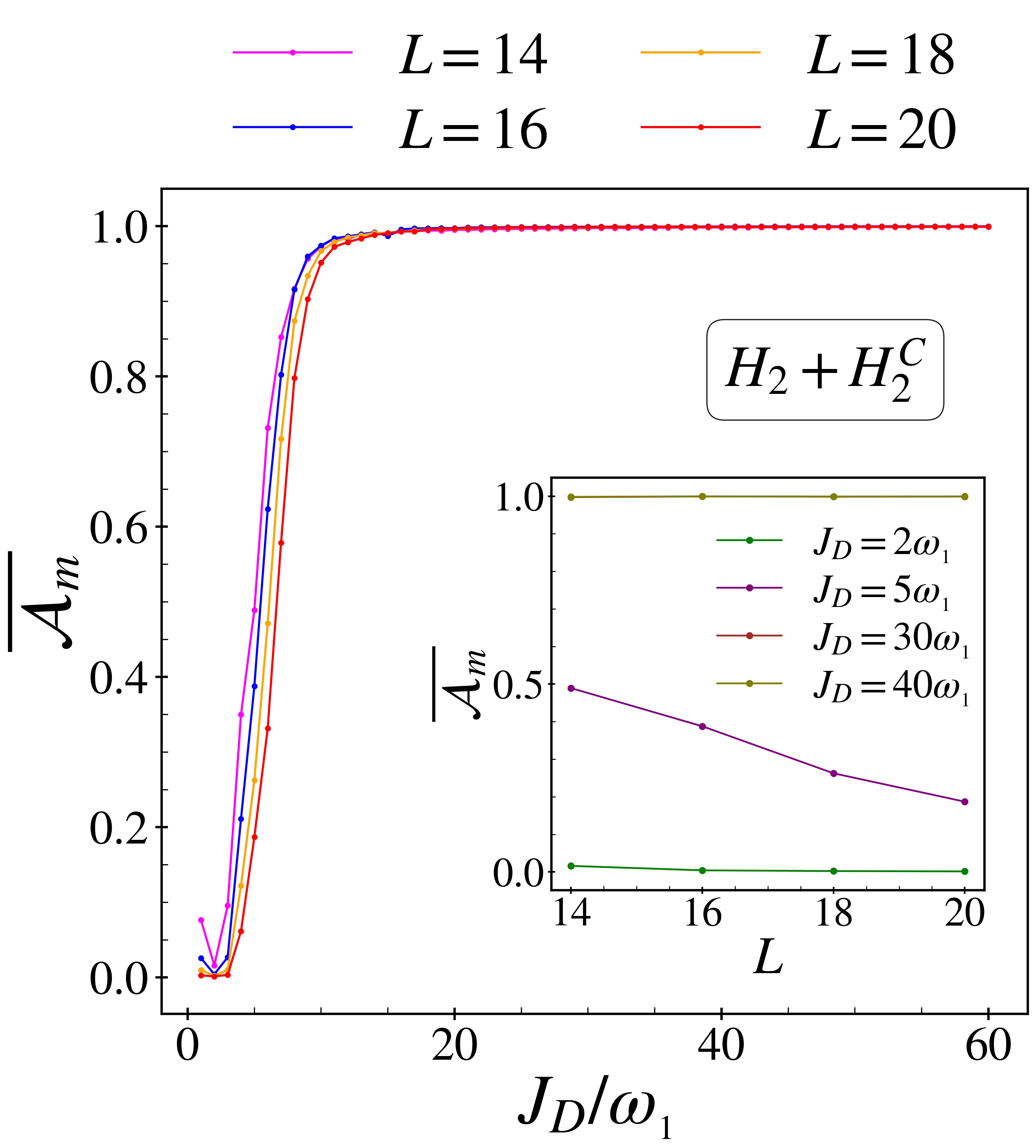}
\caption{
\textbf{Time-independent Hamiltonian engineering for $H_2$.}
\textbf{(Left)} Dynamics under $H_2$ and $H_2+H_2^C$: the
autocorrelation decays for the original Hamiltonian but remains close
to unity after engineering.
\textbf{(Center)} Drive-amplitude scan for $H_2$, showing a systematic
decrease with increasing $L$ throughout the examined range.
\textbf{(Right)} Corresponding result for $H_2+H_2^C$, where this
size dependence disappears beyond a finite drive strength; the inset
shows the $L$ dependence at representative values of $J_D$.
}
\label{fig:SM_static_eng_H2}
\end{figure}

Figure~\ref{fig:SM_static_eng_H2} shows the results for $H_2$. For
the original Hamiltonian, the autocorrelation of $m^x$ decays in time
and its long-time average decreases systematically with increasing
$L$ throughout the examined range of $J_D$. After adding $H_2^C$,
the autocorrelation remains close to unity at strong drive and the
systematic size dependence disappears beyond a finite drive strength.

\begin{figure}[!htbp]
\centering
\includegraphics[width=0.325\textwidth]
{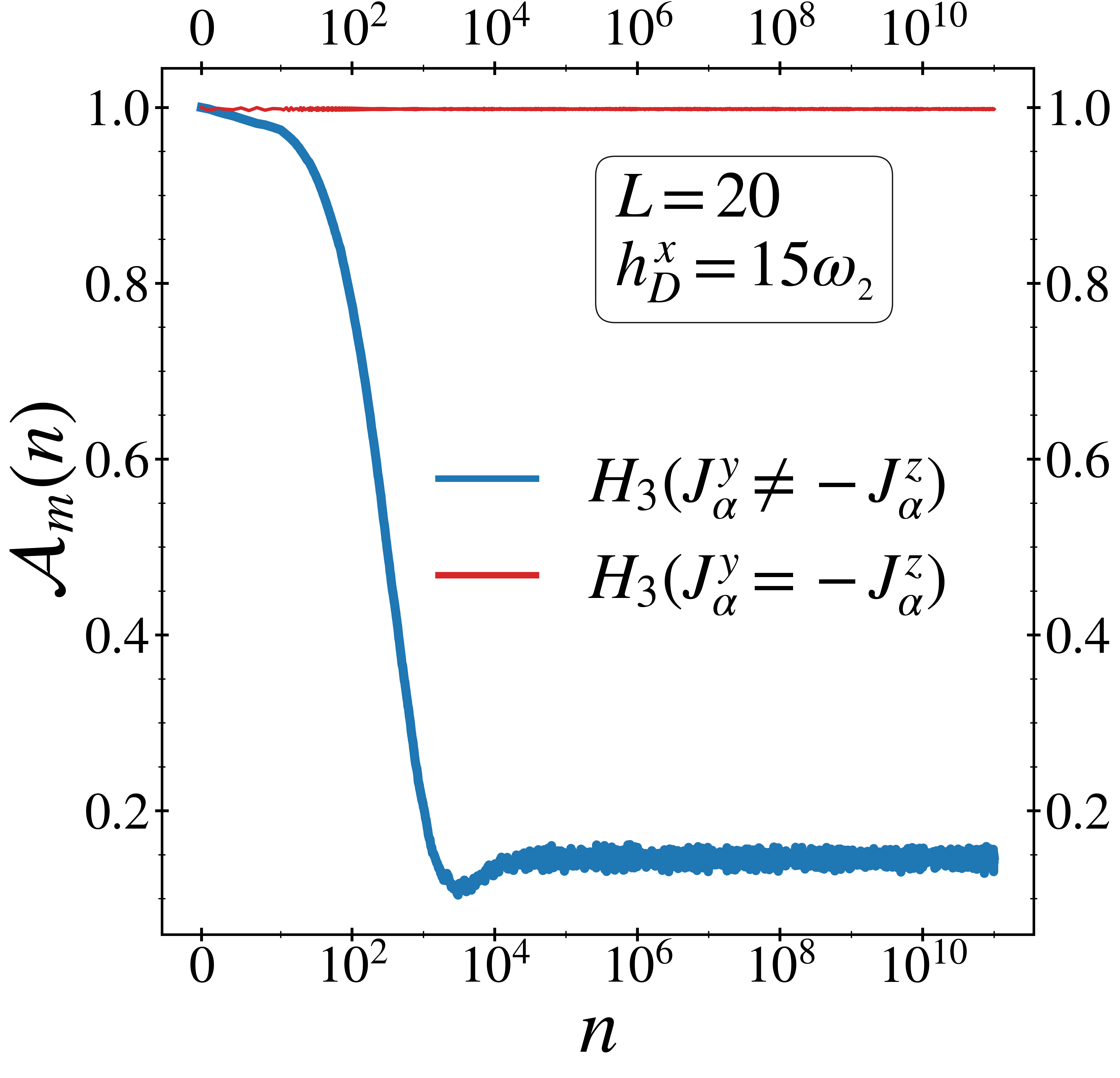}
\hfill
\includegraphics[width=0.325\textwidth]
{Fig_DriveScan_H3_Atcr_mx_2.png}
\hfill
\includegraphics[width=0.325\textwidth]
{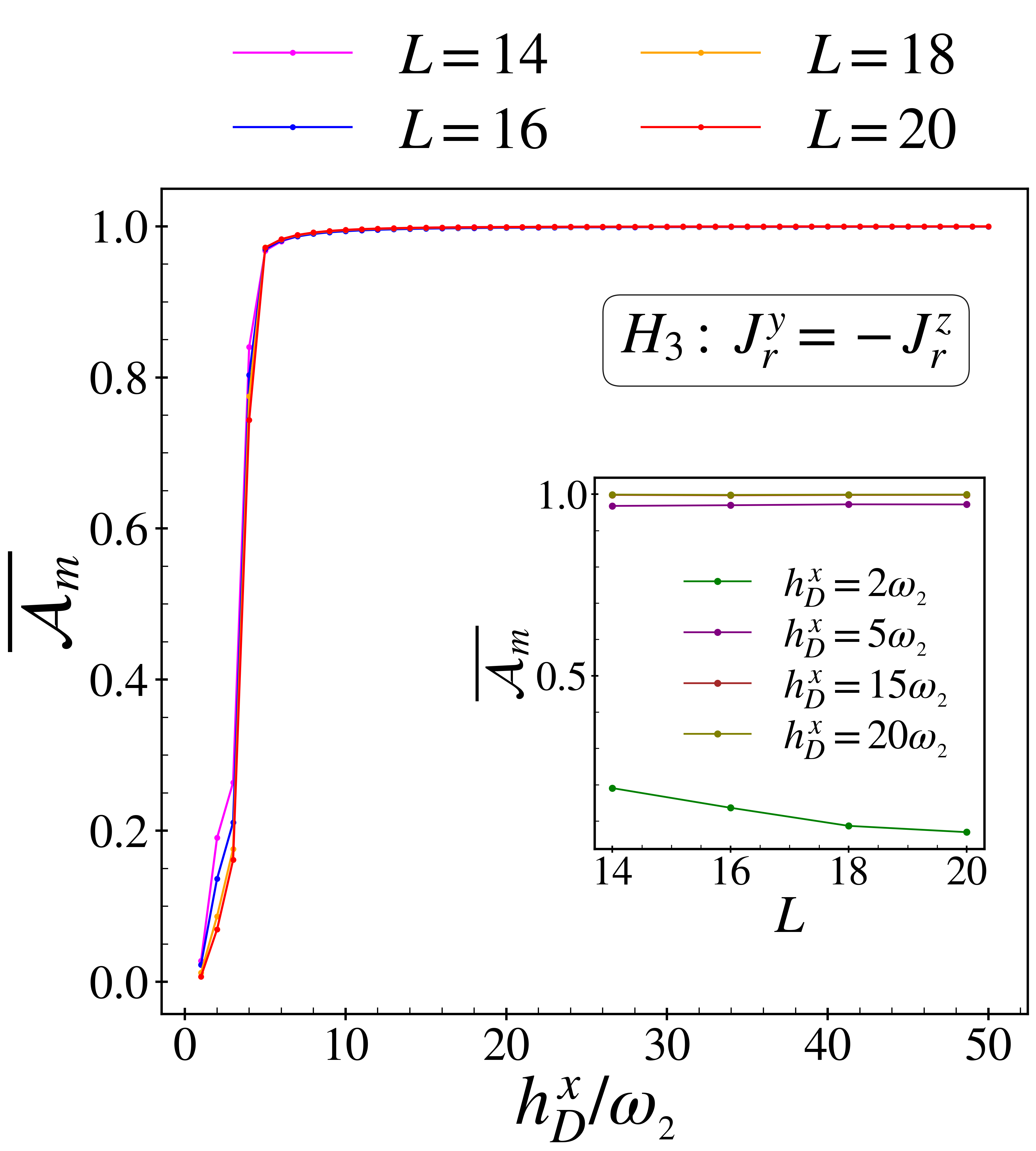}
\caption{
\textbf{Time-independent Hamiltonian engineering for $H_3$.}
\textbf{(Left)} Dynamics before and after imposing
$J_r^y=-J_r^z$: substantial decay occurs in the original model,
whereas the engineered autocorrelation remains close to unity.
\textbf{(Center)} Drive-amplitude scan for the original $H_3$, showing
a systematic decrease with increasing $L$.
\textbf{(Right)} Corresponding engineered result, where this decrease
is restricted to weak drive and disappears beyond a finite drive
strength; the inset shows the $L$ dependence.
}
\label{fig:SM_static_eng_H3}
\end{figure}

The same behavior occurs for $H_3$, as shown in
Fig.~\ref{fig:SM_static_eng_H3}. For the original Hamiltonian, the
autocorrelation shows substantial decay and its long-time average
decreases systematically with increasing $L$. Imposing
$J_r^y=-J_r^z$ $(r=1,2)$ stabilizes the strong-drive dynamics and
removes this systematic size dependence beyond a finite drive
strength.

\begin{figure}[!htbp]
\centering
\includegraphics[width=0.49\textwidth]
{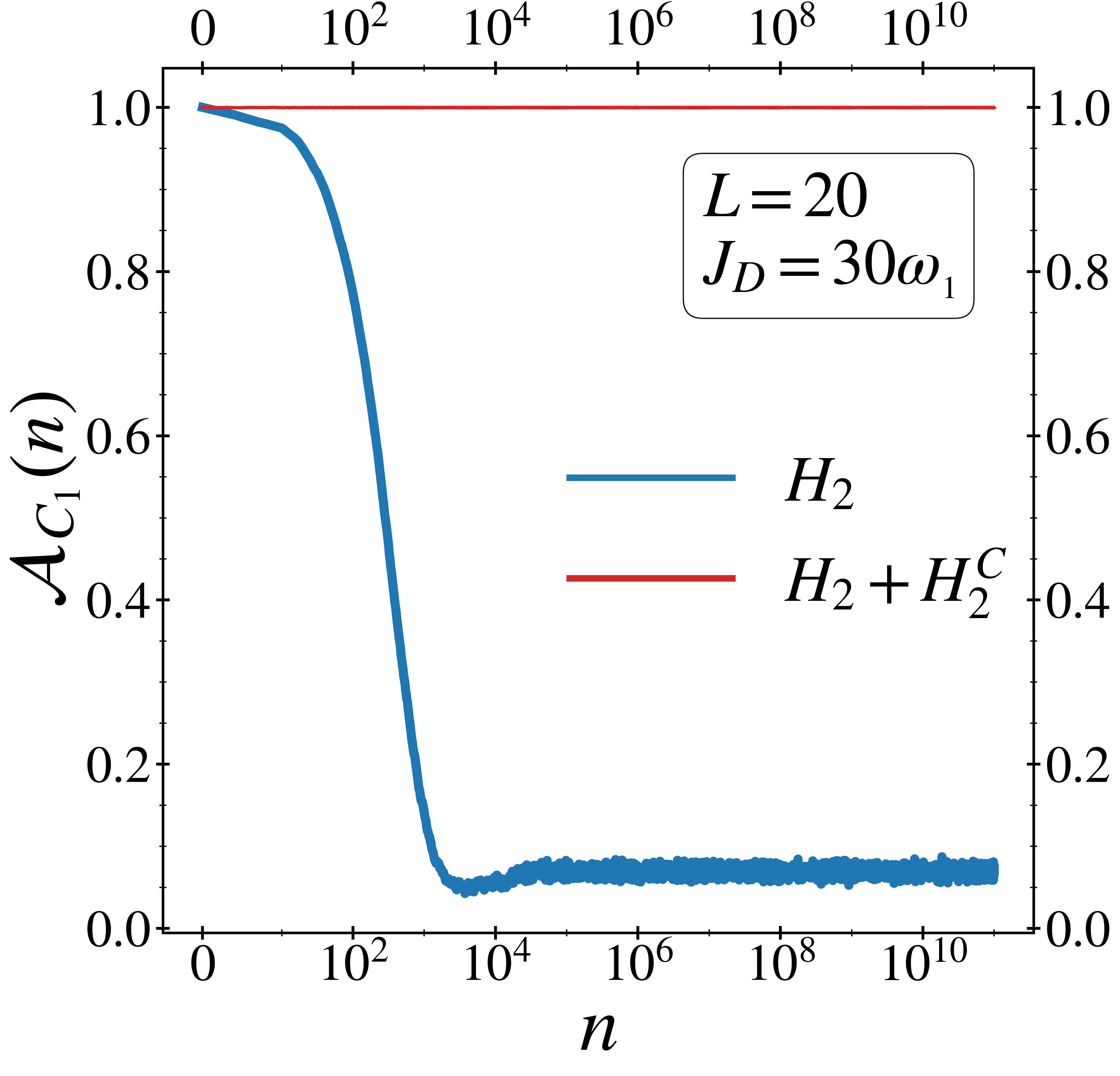}
\hfill
\includegraphics[width=0.49\textwidth]
{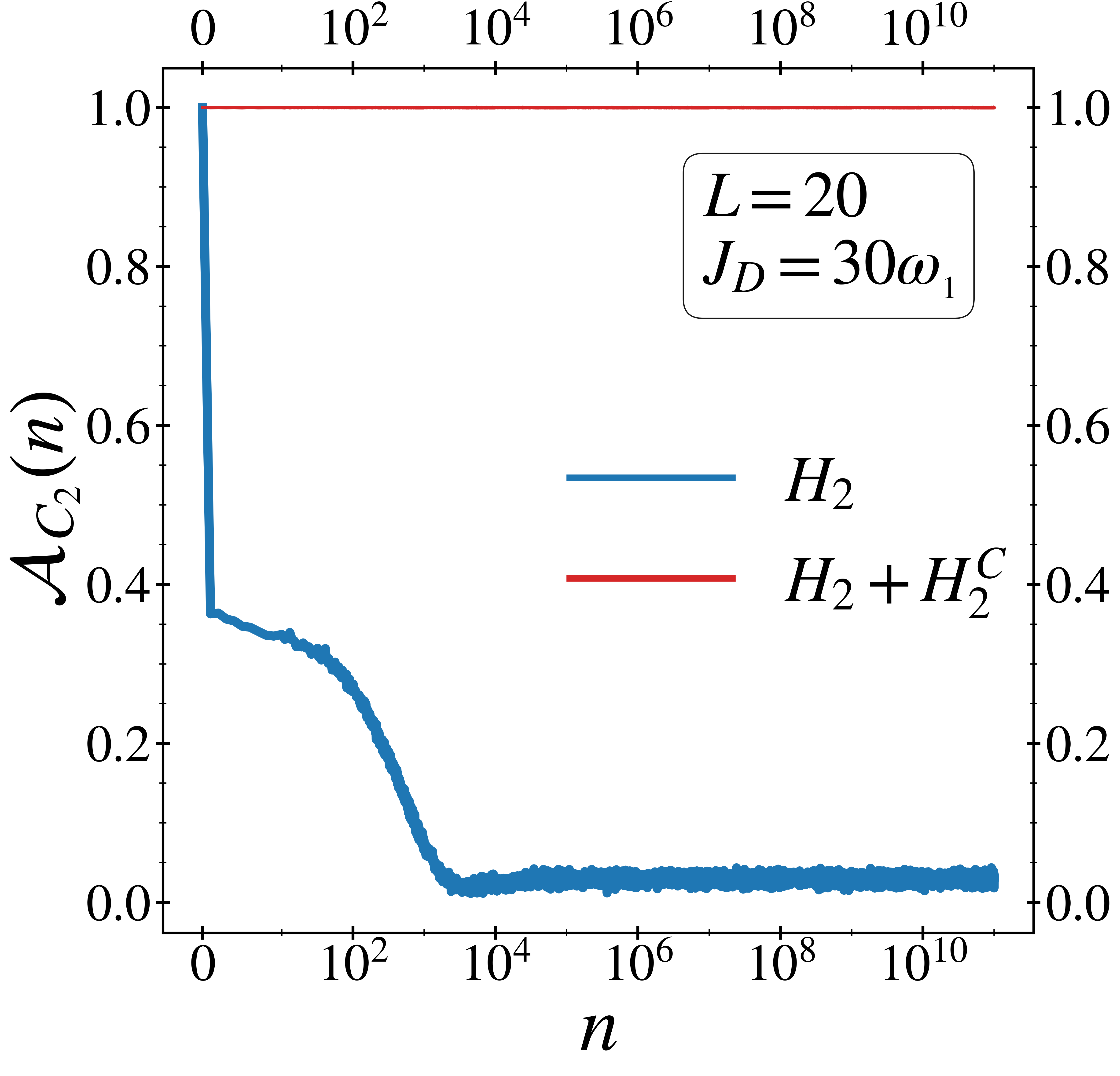}
\caption{
\textbf{Time-independent Hamiltonian engineering of the
autocorrelations of $C_1^x$ and $C_2^x$ for $H_2$.}
\textbf{(Left)} $C_1^x$ and
\textbf{(Right)} $C_2^x$.
For both observables, the autocorrelation decays under $H_2$ but
remains close to unity after adding $H_2^C$.
}
\label{fig:SM_static_eng_H2_C1C2}
\end{figure}

The same engineering of $H_2$ also stabilizes the additional ECOs
$C_1^x$ and $C_2^x$, as shown in
Fig.~\ref{fig:SM_static_eng_H2_C1C2}.

\subsection{Drive-operator engineering:
Real-time dynamics, amplitude scans, and system-size dependence}

We next present additional numerical results for the drive-engineering
route, in which $H_{\rm stat}$ is unchanged while the drive operator
is modified so that the relevant quantum terms are transferred from
the $q=0$ sector to the $q \neq 0$ sectors.

\begin{figure}[!htbp]
\centering
\includegraphics[width=0.48\textwidth]
{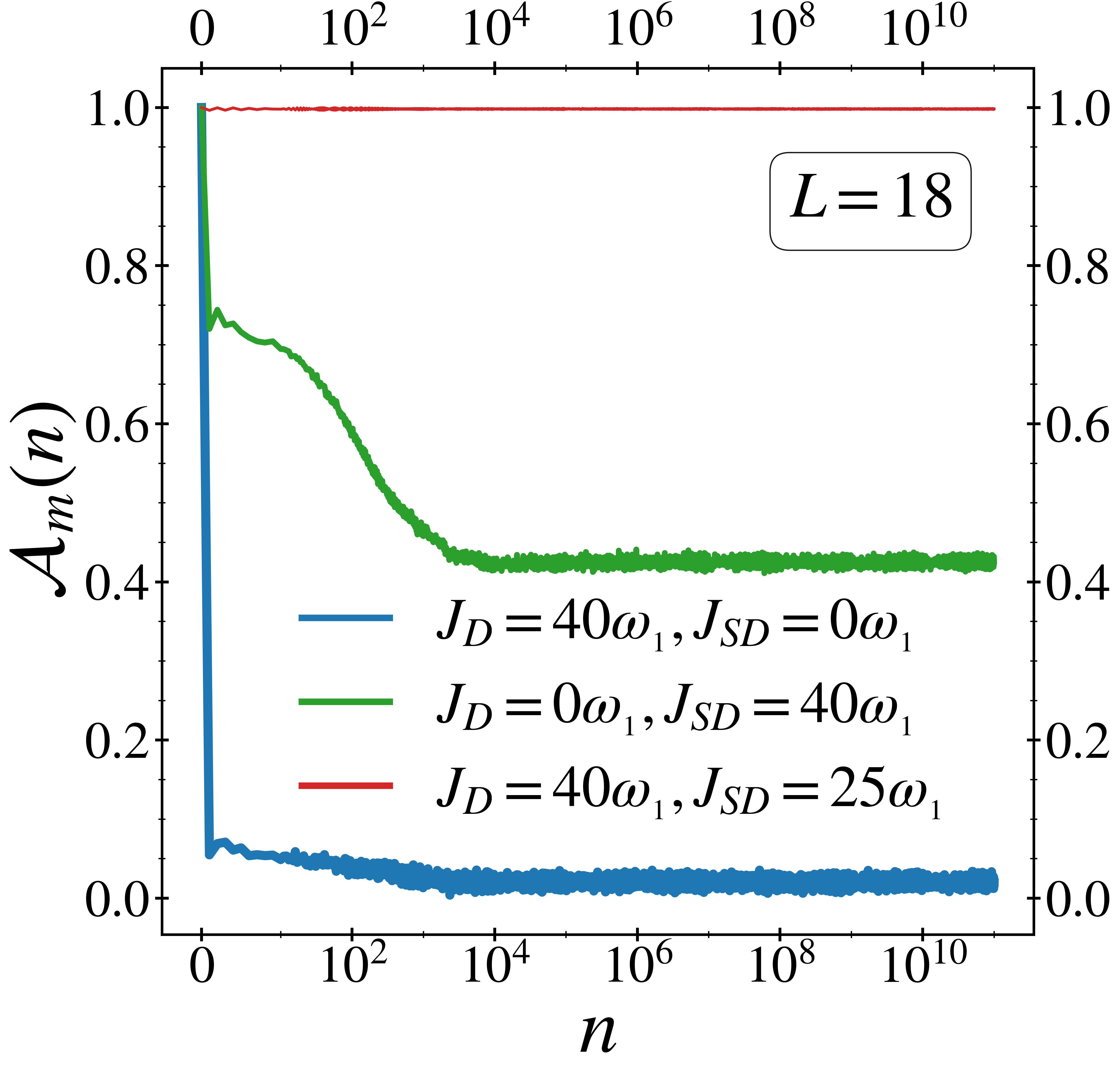}
\hfill
\includegraphics[width=0.48\textwidth]
{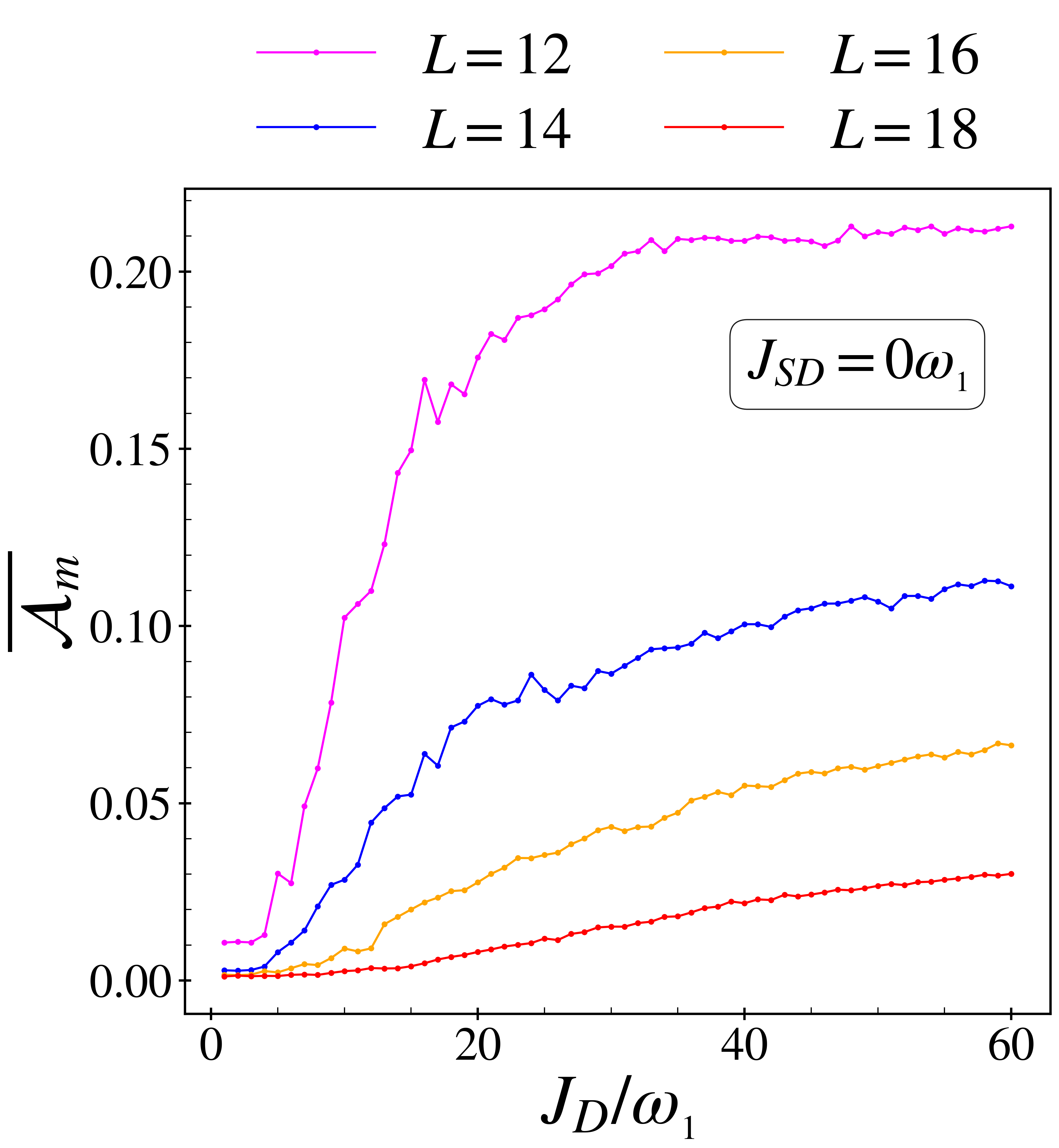}

\vspace{0.5em}

\includegraphics[width=0.48\textwidth]
{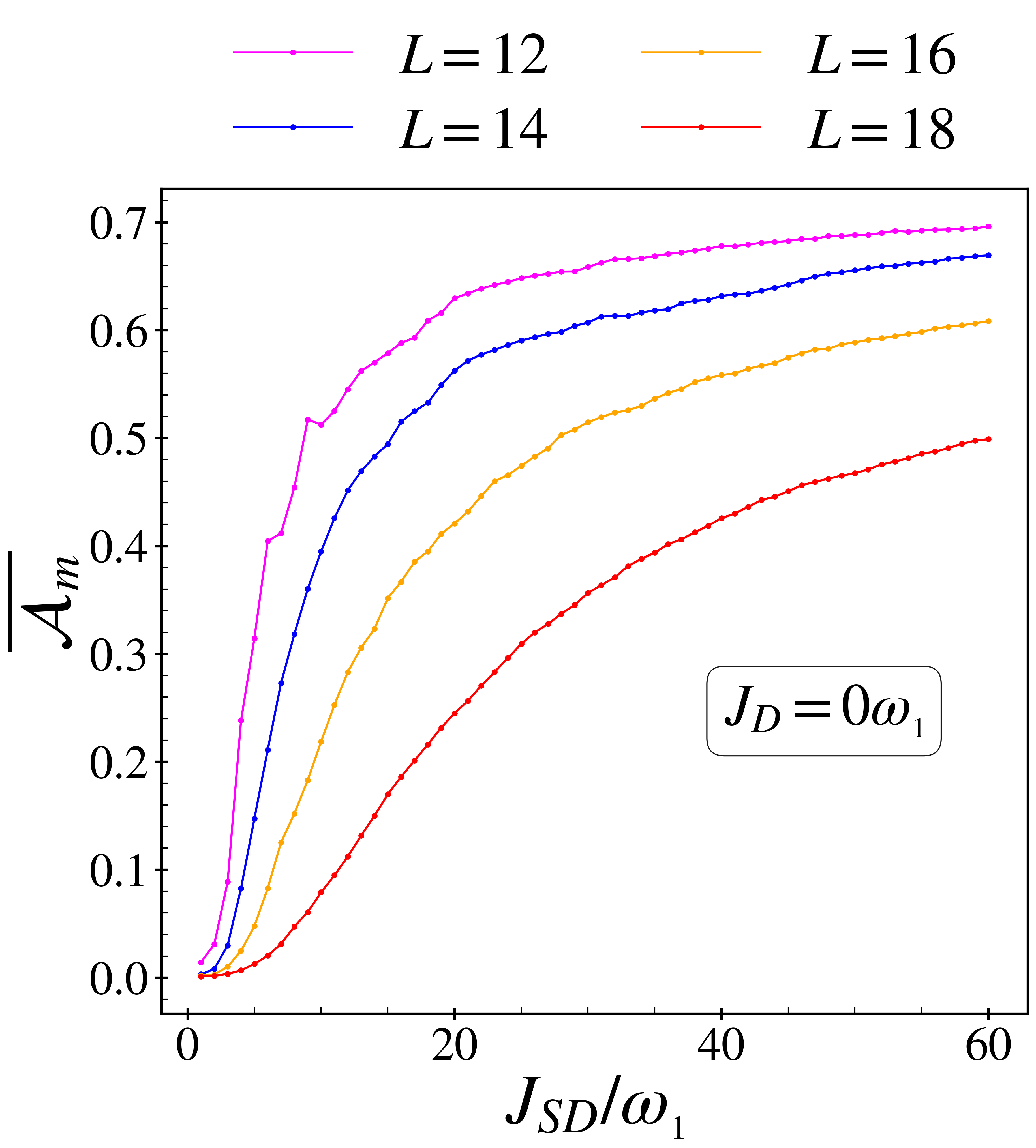}
\hfill
\includegraphics[width=0.48\textwidth]
{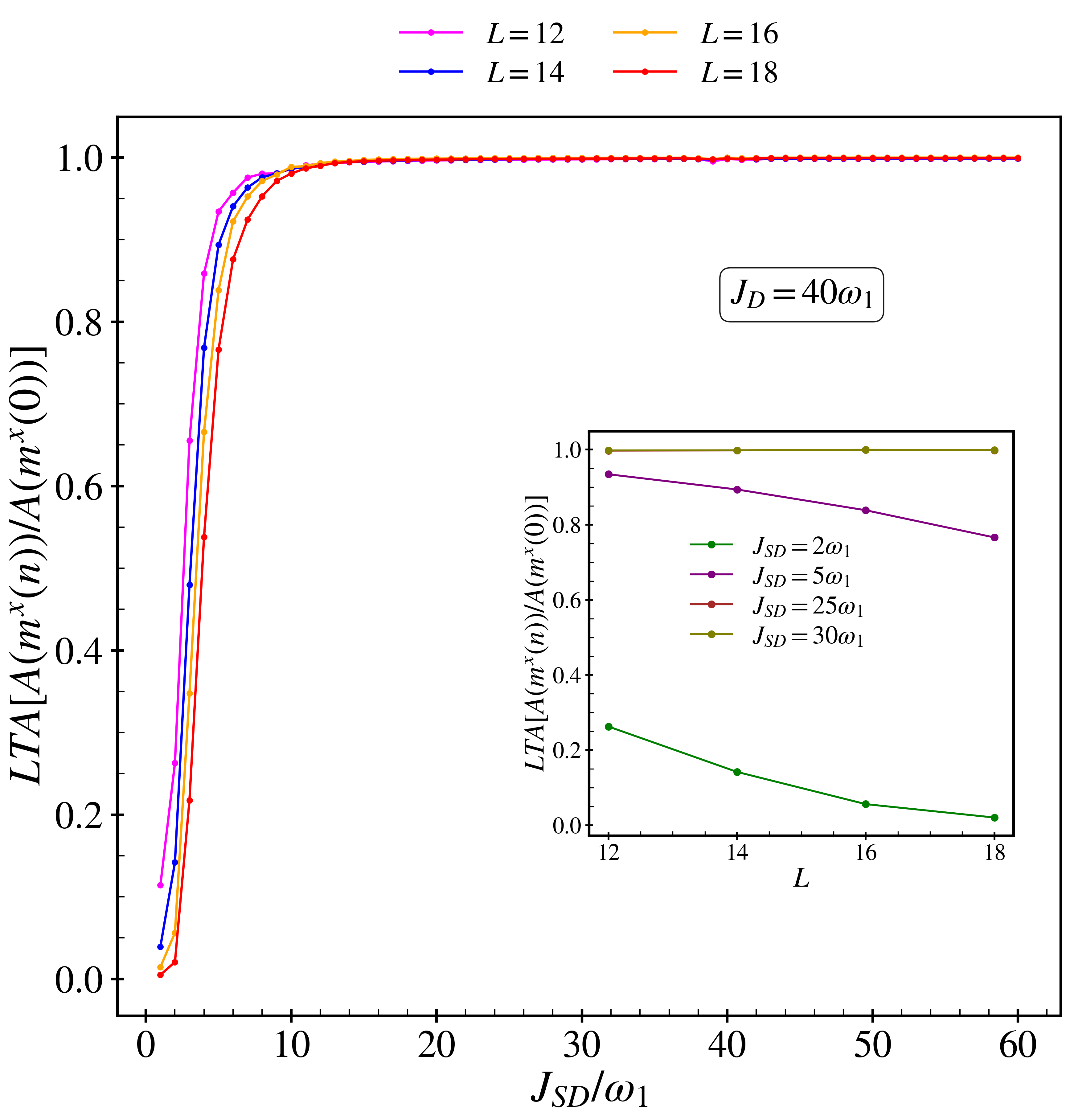}
\caption{
\textbf{Drive-operator engineering for $H_2$.}
\textbf{(Top left)} Dynamics for a uniform bond drive
($J_D=40\omega_{_1}$, $J_{SD}=0$), a staggered bond drive
($J_D=0$, $J_{SD}=40\omega_{_1}$), and the engineered drive
($J_D=40\omega_{_1}$, $J_{SD}=25\omega_{_1}$); persistent freezing occurs
only for the combined drive.
\textbf{(Top right)} $J_D$ scan for $J_{SD}=0$.
\textbf{(Bottom left)} $J_{SD}$ scan for $J_D=0$.
Both show a systematic decrease with increasing $L$.
\textbf{(Bottom right)} $J_{SD}$ scan at fixed
$J_D=40\omega_{_1}$, where this size dependence disappears beyond a
finite $J_{SD}$; the inset shows the corresponding $L$ dependence.
}
\label{fig:SM_drive_eng_H2}
\end{figure}

For $H_2$, neither the uniform bond drive ($J_{SD}=0$) nor the
staggered bond drive ($J_D=0$) alone produces stable DF. When both
components of $D_2^x$ are present, however, the strong-drive
autocorrelation remains close to unity and the systematic size
dependence disappears, as shown in Fig.~\ref{fig:SM_drive_eng_H2}.

\begin{figure}[!htbp]
\centering
\includegraphics[width=0.48\textwidth]
{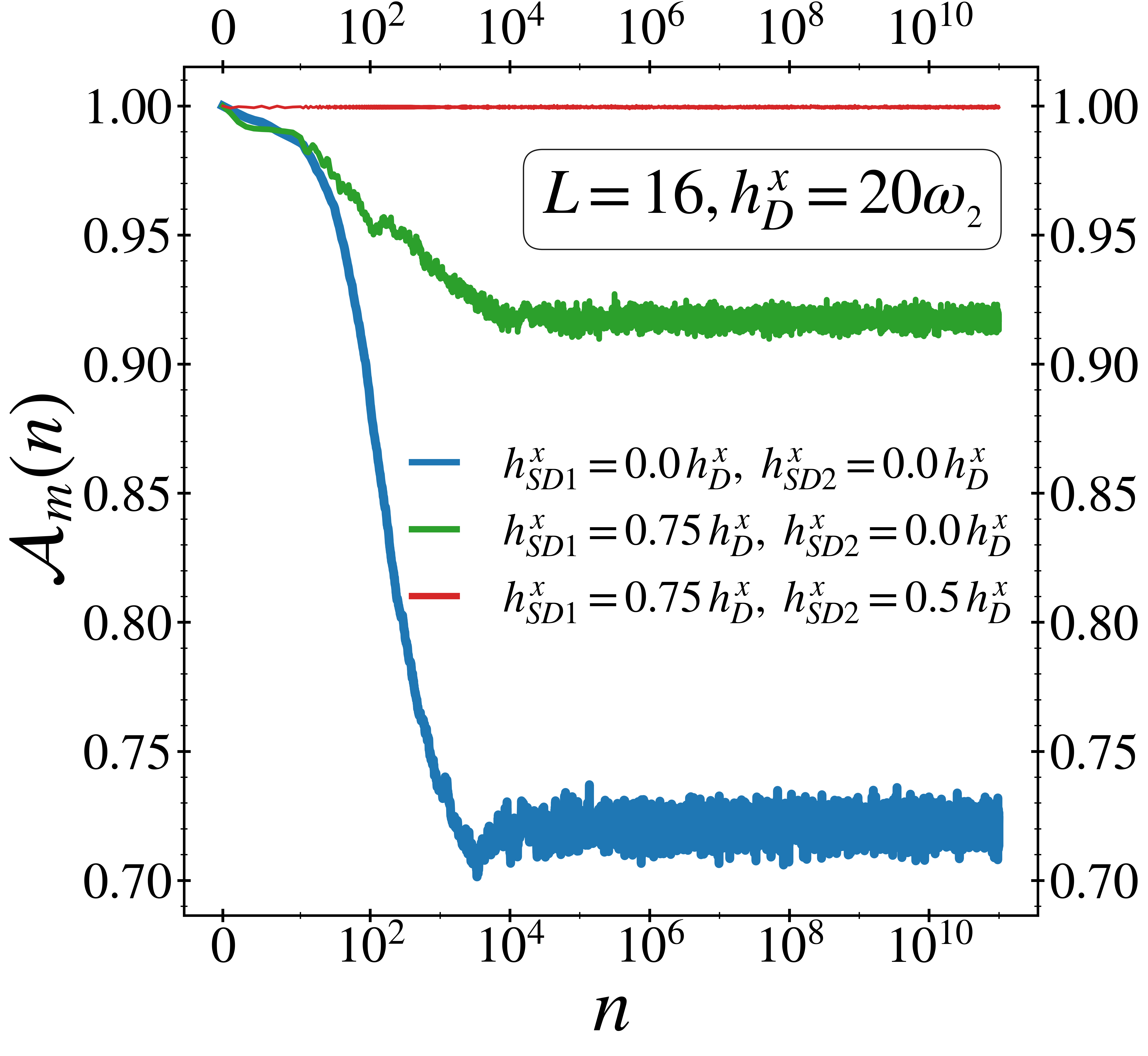}
\hfill
\includegraphics[width=0.48\textwidth]
{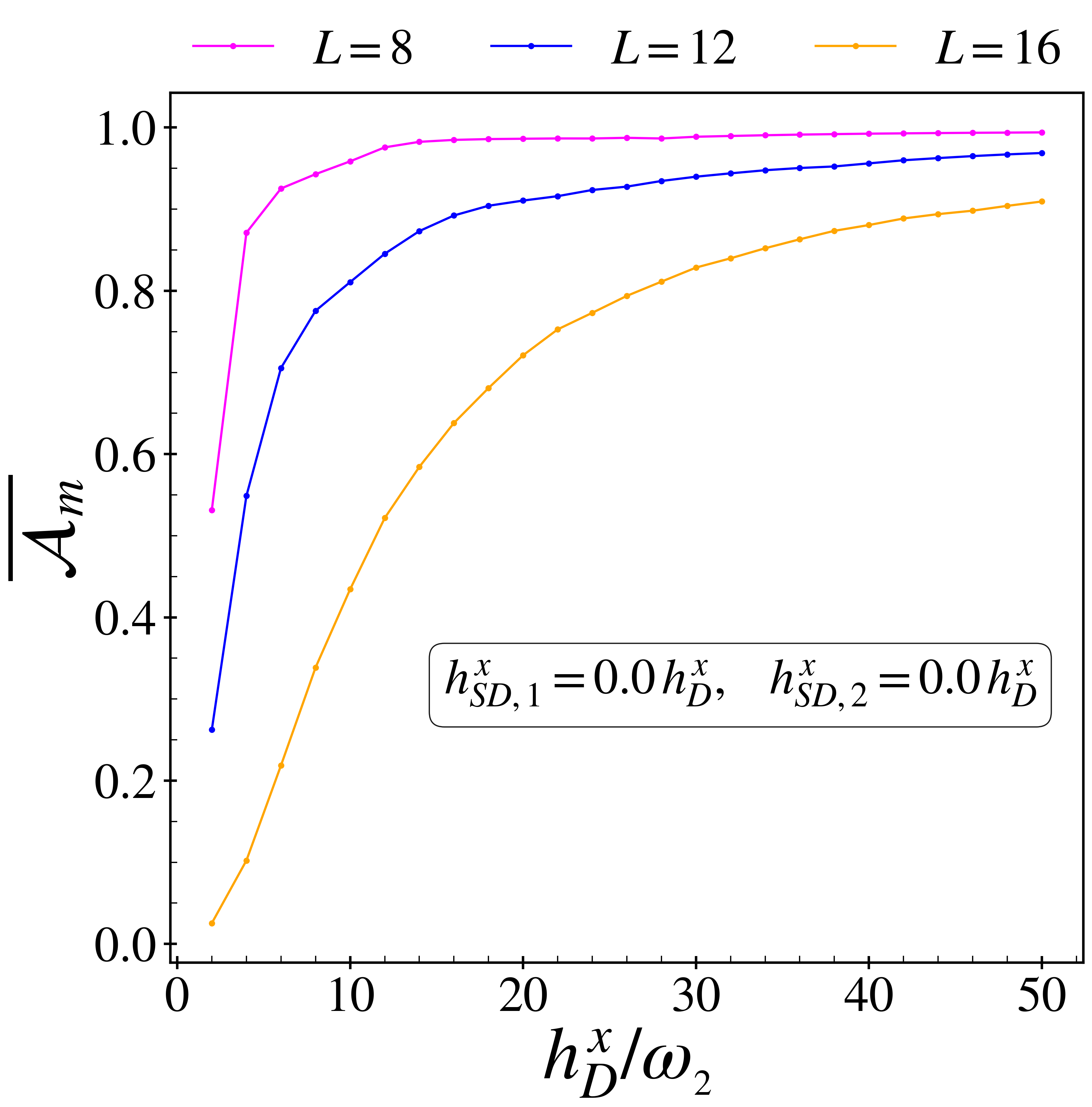}

\vspace{0.5em}

\includegraphics[width=0.48\textwidth]
{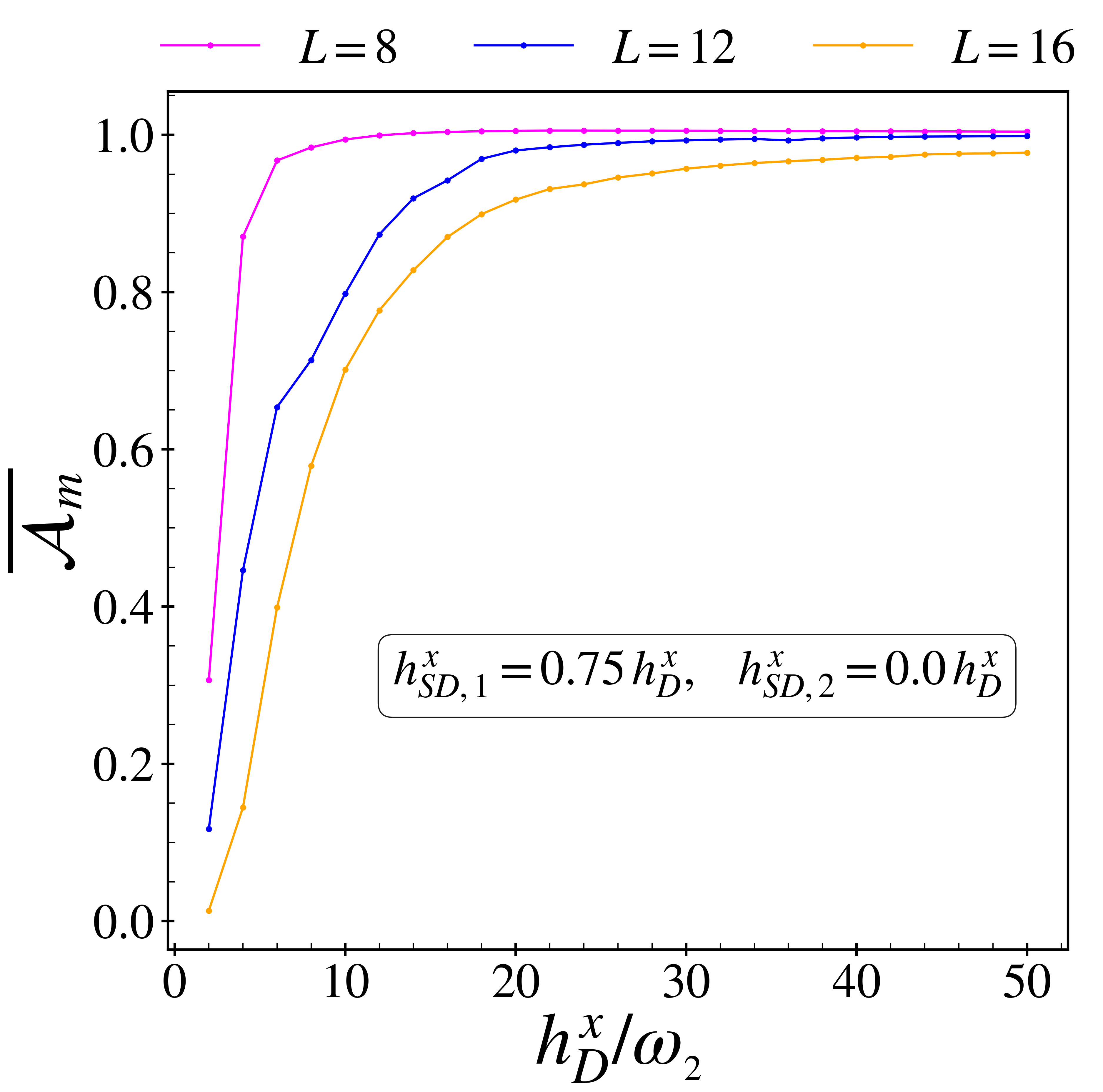}
\hfill
\includegraphics[width=0.48\textwidth]
{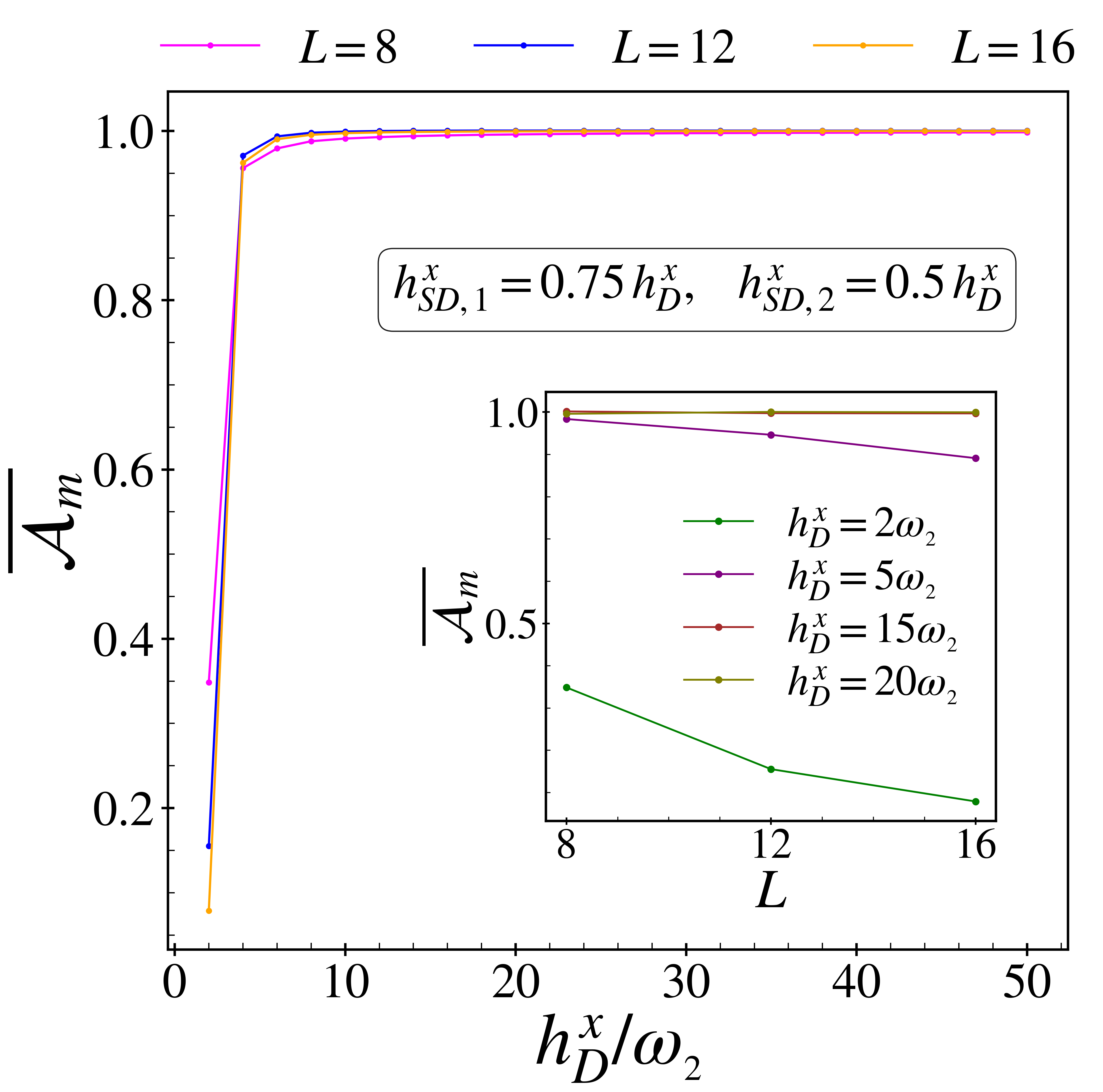}
\caption{
\textbf{Drive-operator engineering for $H_3$.}
\textbf{(Top left)} Dynamics for the uniform field drive
($h_{SD,1}^x=h_{SD,2}^x=0$), after adding only the first staggered
component ($h_{SD,1}^x=0.75h_D^x$, $h_{SD,2}^x=0$), and after adding
both components ($h_{SD,1}^x=0.75h_D^x$,
$h_{SD,2}^x=0.5h_D^x$).
\textbf{(Top right)} Drive-amplitude scan for the uniform drive,
showing a systematic decrease with increasing $L$.
\textbf{(Bottom left)} Result with only the first staggered component:
the retained autocorrelation is enhanced, but the size dependence
persists.
\textbf{(Bottom right)} Fully engineered $D_3^x$, where the size
dependence disappears beyond a finite drive strength; the inset shows
the corresponding $L$ dependence.
}
\label{fig:SM_drive_eng_H3}
\end{figure}

Figure~\ref{fig:SM_drive_eng_H3} shows the corresponding results for
$H_3$. The uniform field drive alone does not stabilize the
autocorrelation. Adding the first staggered component,
$h_{SD,1}^x=0.75h_D^x$, substantially enhances the retained memory
but does not remove its systematic size dependence. Including the
second component, $h_{SD,2}^x=0.5h_D^x$, stabilizes the strong-drive
autocorrelation and removes this size dependence beyond a finite
drive strength.

\begin{figure}[!htbp]
\centering
\includegraphics[width=0.49\textwidth]
{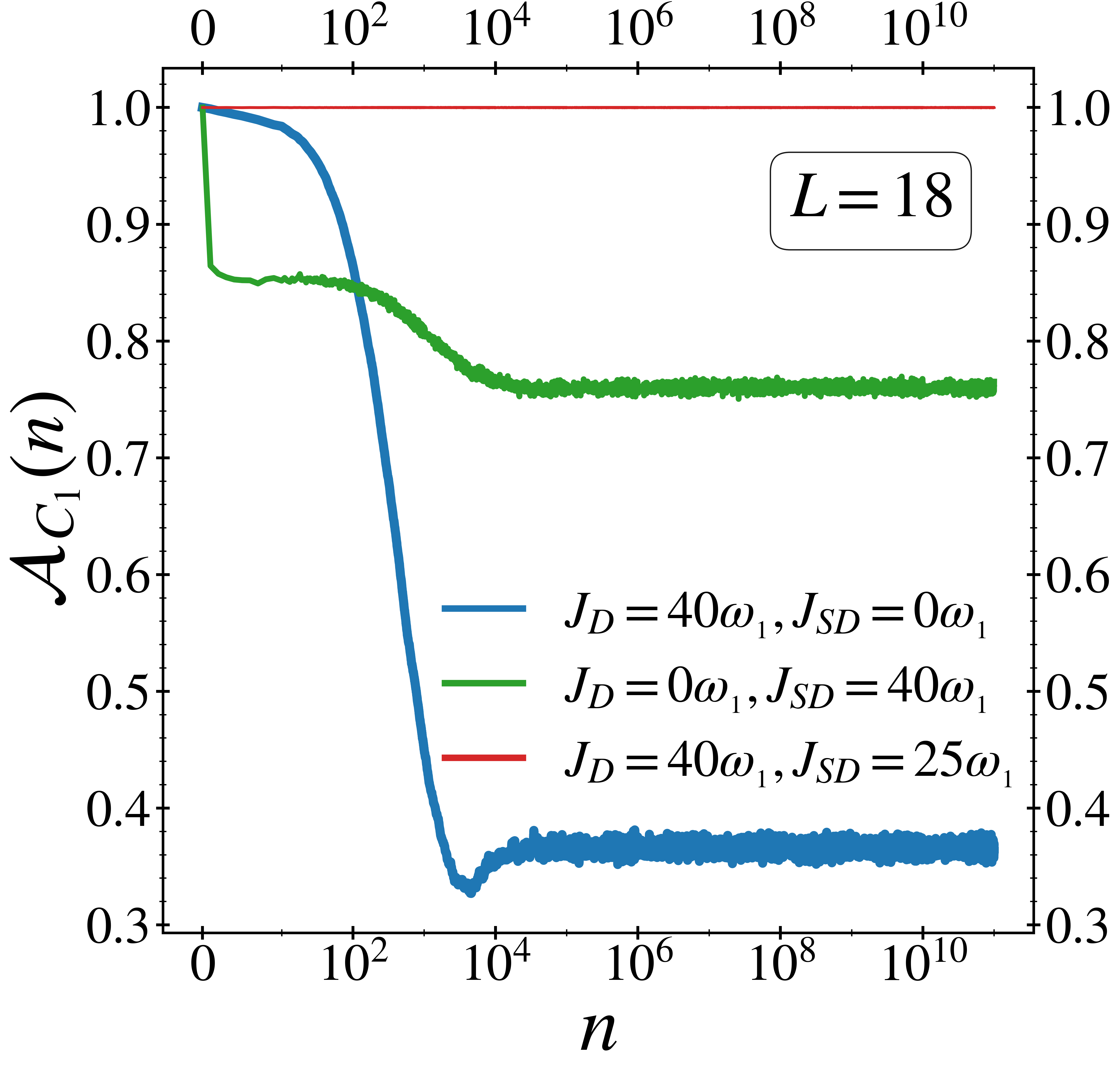}
\hfill
\includegraphics[width=0.49\textwidth]
{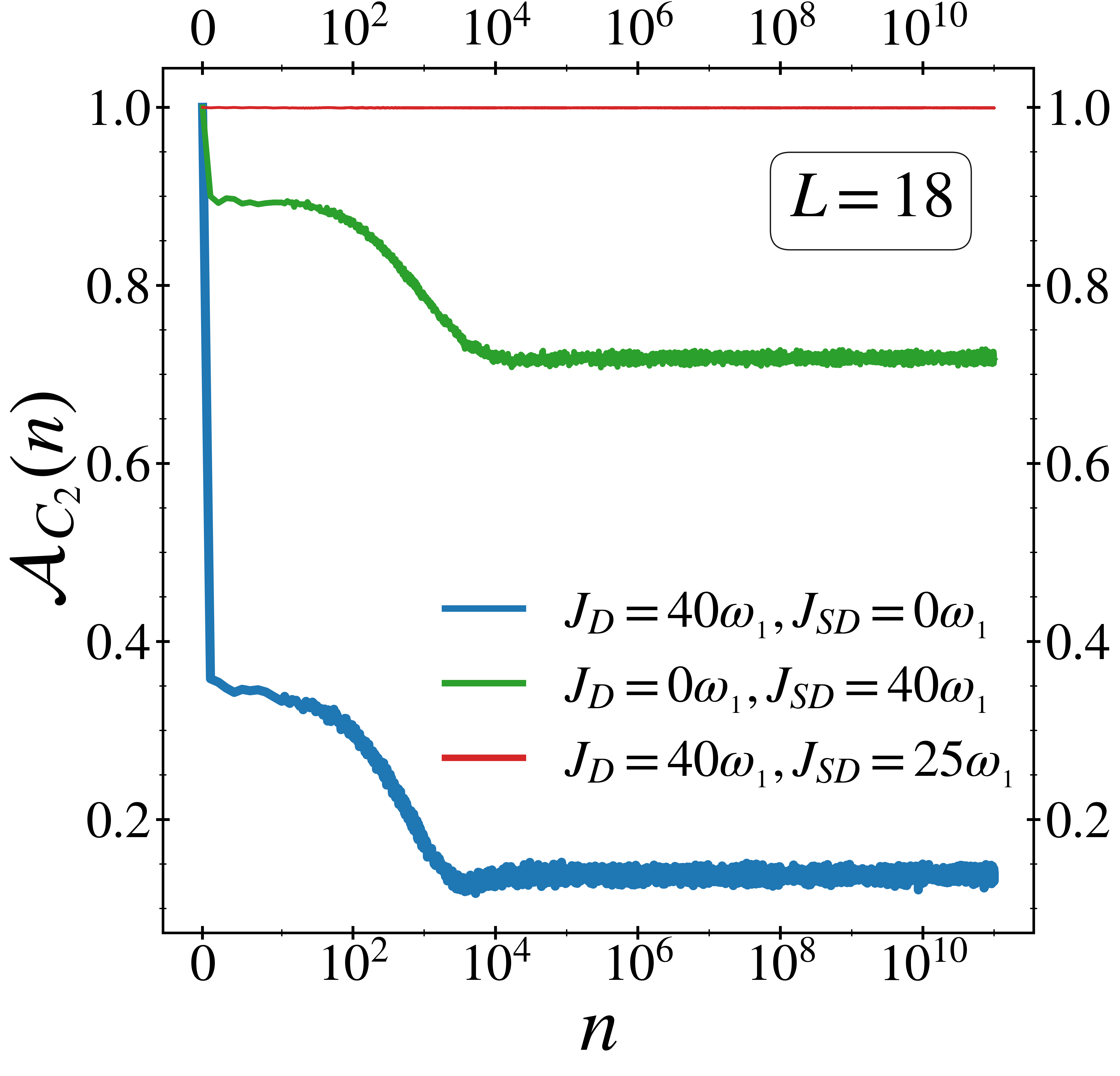}
\caption{
\textbf{Drive-operator engineering of the autocorrelations of
$C_1^x$ and $C_2^x$ for $H_2$.}
\textbf{(Left)} $C_1^x$ and
\textbf{(Right)} $C_2^x$.
For both observables, the uniform bond drive
($J_D=40\omega_{_1}$, $J_{SD}=0$) and staggered bond drive
($J_D=0$, $J_{SD}=40\omega_{_1}$) individually show substantial loss of
memory, whereas the engineered drive
($J_D=40\omega_{_1}$, $J_{SD}=25\omega_{_1}$) retains an autocorrelation
close to unity throughout the shown time window.
}
\label{fig:SM_drive_eng_H2_C1C2}
\end{figure}

Drive-operator engineering of $H_2$ also stabilizes the additional
ECOs $C_1^x$ and $C_2^x$, as shown in
Fig.~\ref{fig:SM_drive_eng_H2_C1C2}.

\subsection{Numerical evidence for infinite-time ECO stability from thermal mixed-state DEAs}

We further test the engineered ECOs using the diagonal-ensemble
average (DEA) of a high-temperature thermal mixed state with
$\beta=0.05$. The DEA gives the infinite-time stroboscopic average of
an observable and therefore probes the persistence of its memory at
arbitrarily late times. The high-temperature mixed state has support
over a broad portion of the many-body spectrum.

\begin{figure}[!htbp]
\centering
\includegraphics[width=0.45\textwidth]
{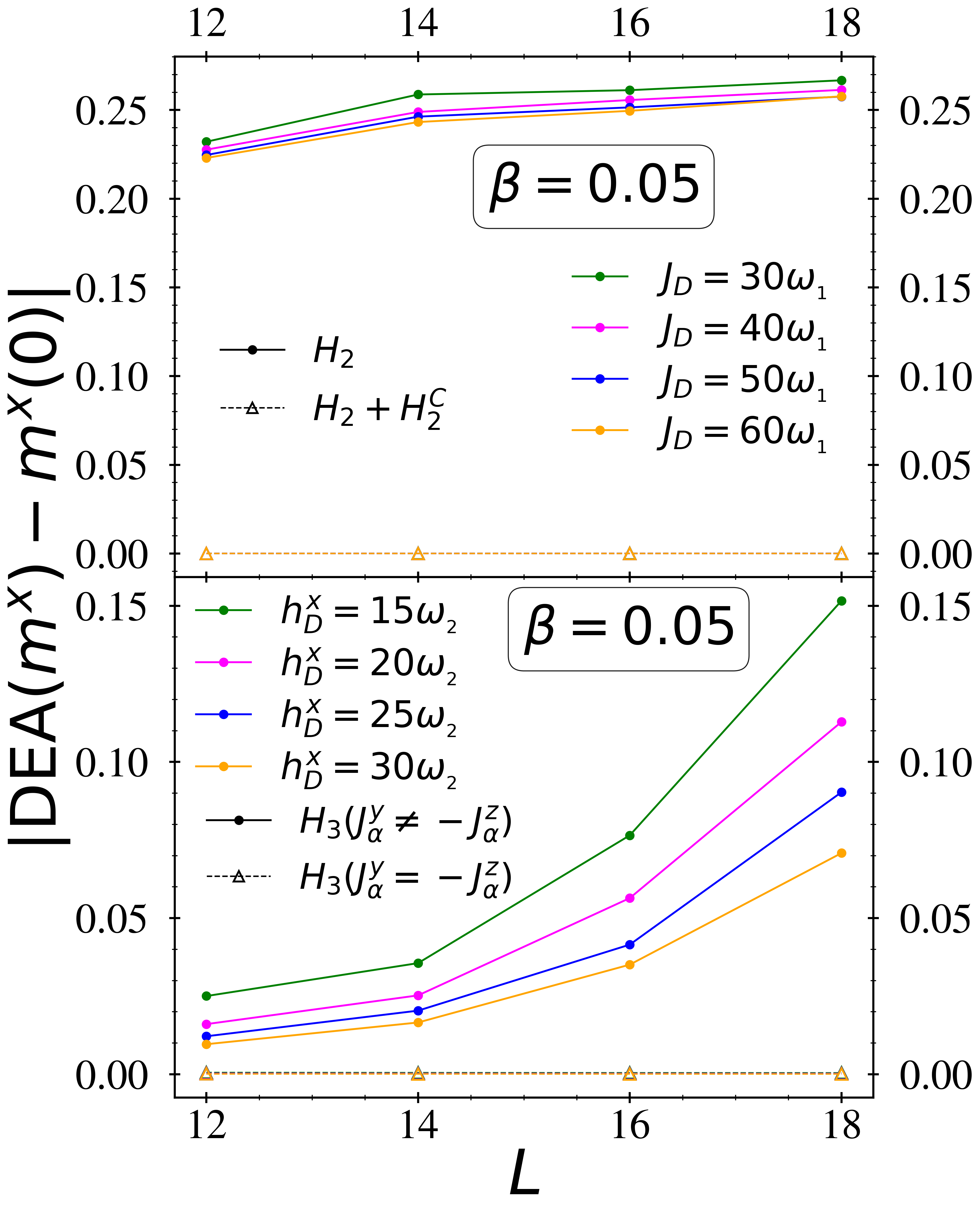}
\hfill
\includegraphics[width=0.45\textwidth]
{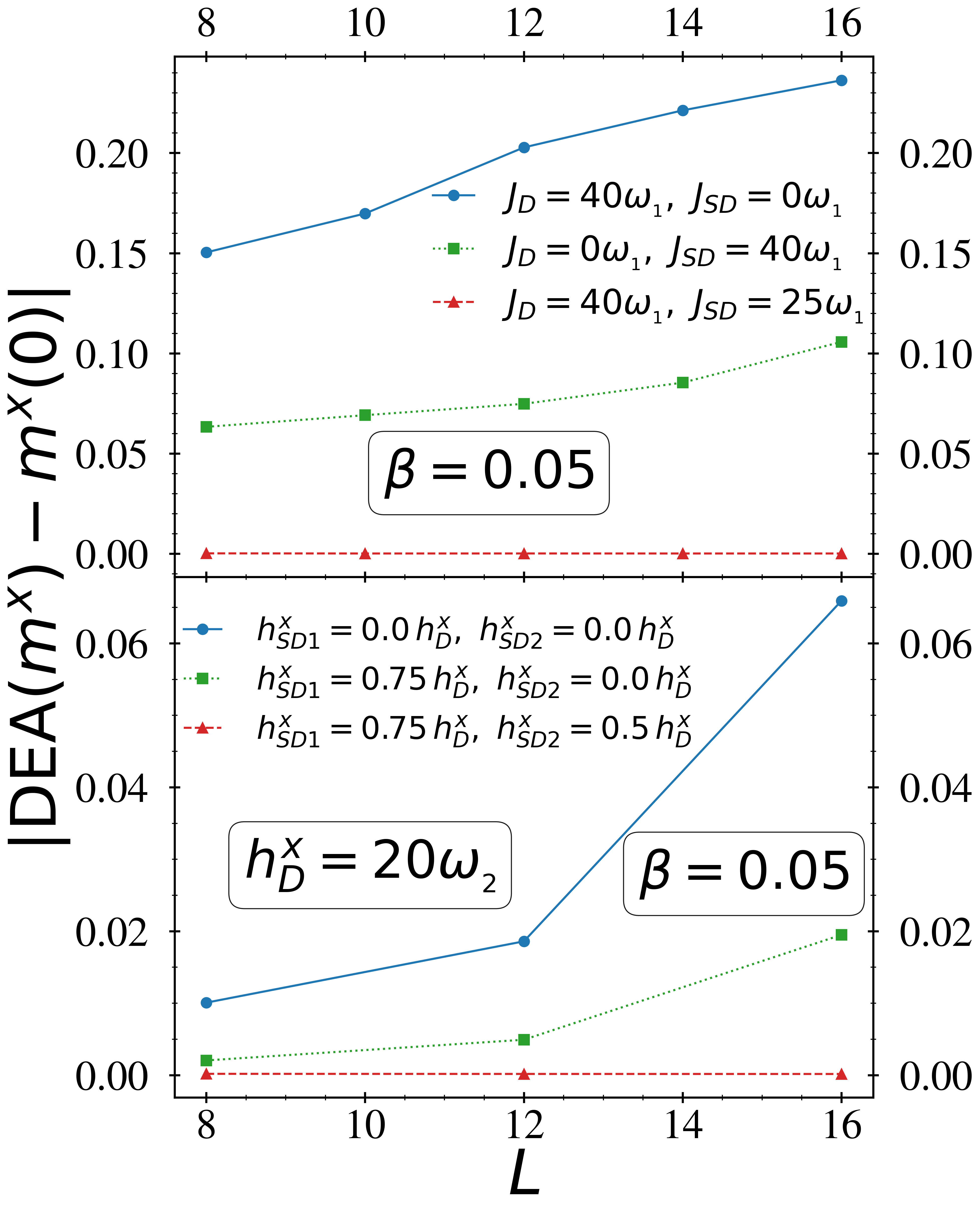}
\caption{
\textbf{Infinite-time memory of $m^x$ from the DEA of a thermal
mixed state with $\beta=0.05$.}
The memory is quantified by
$|\mathrm{DEA}(m^x)-m^x(0)|$.
\textbf{(Left)} Time-independent Hamiltonian engineering and
\textbf{(Right)} drive-operator engineering, with $H_2$ and $H_3$
shown in the top and bottom panels, respectively. For the original
Hamiltonians, the deviation increases systematically with $L$; this
trend is absent for the engineered models, for which the DEA remains
close to the initial value.
}
\label{fig:SM_DEA_engineering}
\end{figure}

Figure~\ref{fig:SM_DEA_engineering} shows
$|\mathrm{DEA}(m^x)-m^x(0)|$ as a function of system size. For the
original Hamiltonians, the deviation increases systematically with
$L$, whereas this trend is absent after either time-independent
Hamiltonian engineering or drive-operator engineering.

\subsection{Spin-flip Ordered Series}

We now examine the spin-flip ordered series of the effective Floquet Hamiltonian introduced in the main text. In the $x$-basis, the effective Hamiltonian is decomposed according to the Hamming distance $\xi$ between pairs of $x$-bit-string configurations,
as defined in Eq.~(15) of the main text. The weight of each
spin-flip sector is quantified by its Frobenius norm,
\begin{equation}
    \left\|H_{\mathrm{eff}}^{(\xi)}\right\|
    =
    \sqrt{
    \sum_{\substack{i,j\\
    \mathrm{dist}(i,j)=\xi}}
    \left|H_{\mathrm{eff}}^{i,j}\right|^2
    }.
\end{equation}

We first consider the field-driven Ising model $H_1(t)$, which already
hosts DF and therefore provides a direct contrast between weak and strong driving. The top-left panel of Fig.~S11 shows $\|H_{\rm eff}^{(\xi)}\|$ for $h_D^x=2\omega_1$ and $h_D^x=30\omega_1$, for $L=8,10,12,14$. At the weaker drive, substantial weight extends to higher spin-flip orders, with the profile broadening as the system size is increased. The dominant off-diagonal weight is concentrated around $\xi \approx L/2$. In contrast, at $h_D^x=30\omega_{_1}$, the weight is strongly concentrated in the lowest spin-flip sectors and rapidly decreases with increasing $\xi$. The onset of this decay occurs at approximately the same $\xi$ for the different system sizes, consistent with the quasi-local structure associated with the dynamically frozen regime.\\

\begin{figure*}[t]
\centering

\includegraphics[width=0.47\textwidth]
{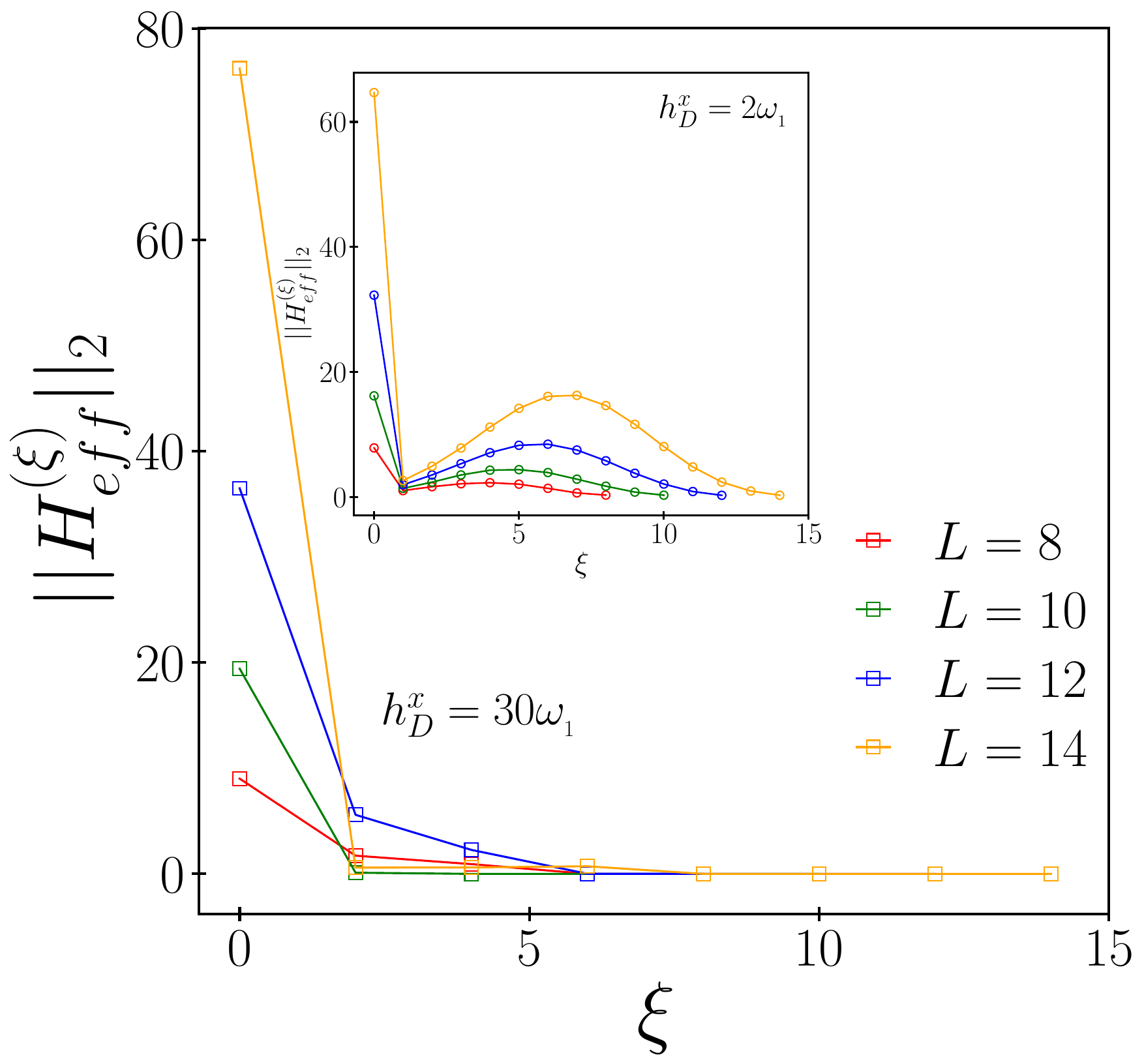}
\hfill
\includegraphics[width=0.47\textwidth]
{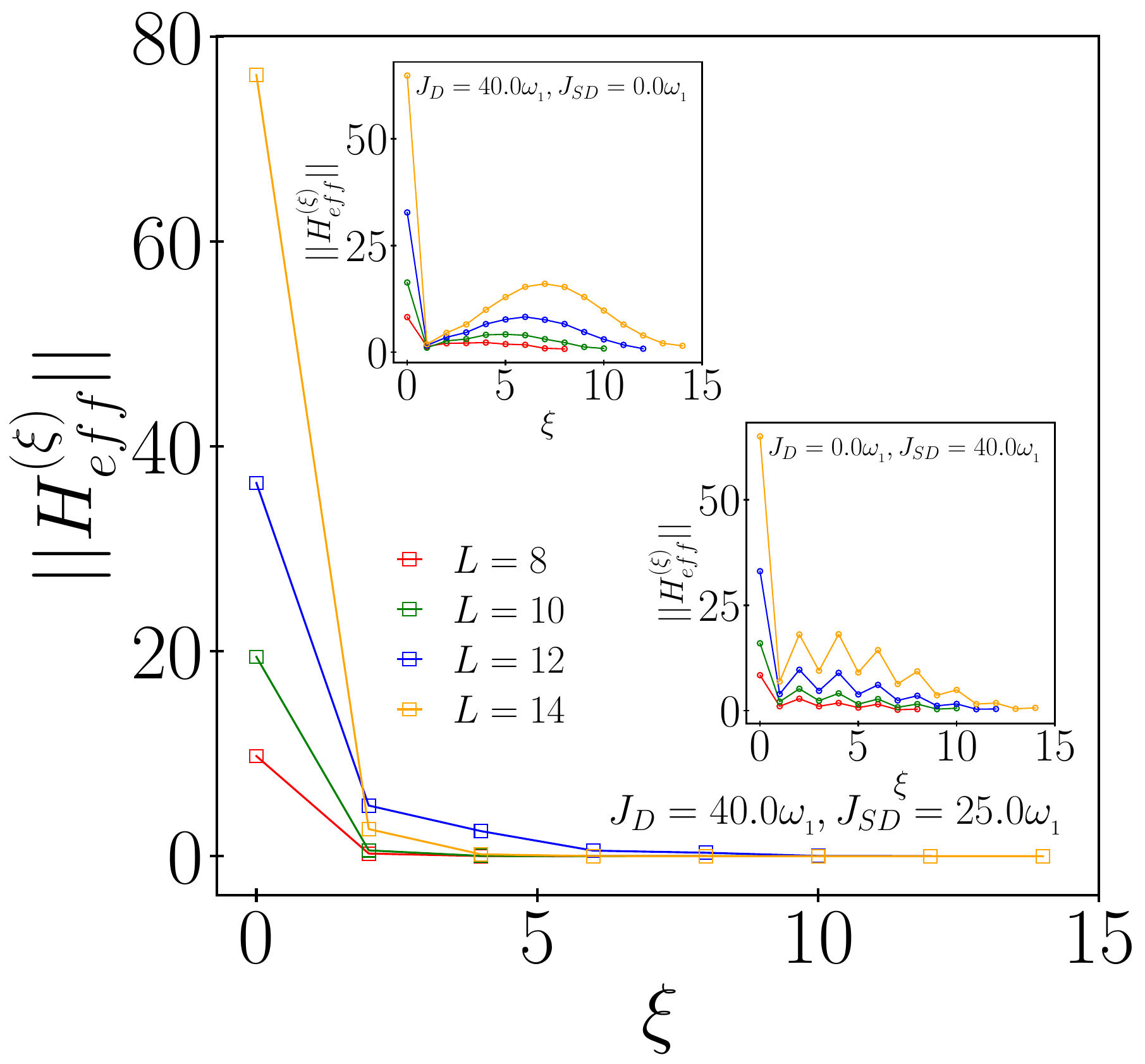}

\vspace{0.5em}

\includegraphics[width=0.47\textwidth]
{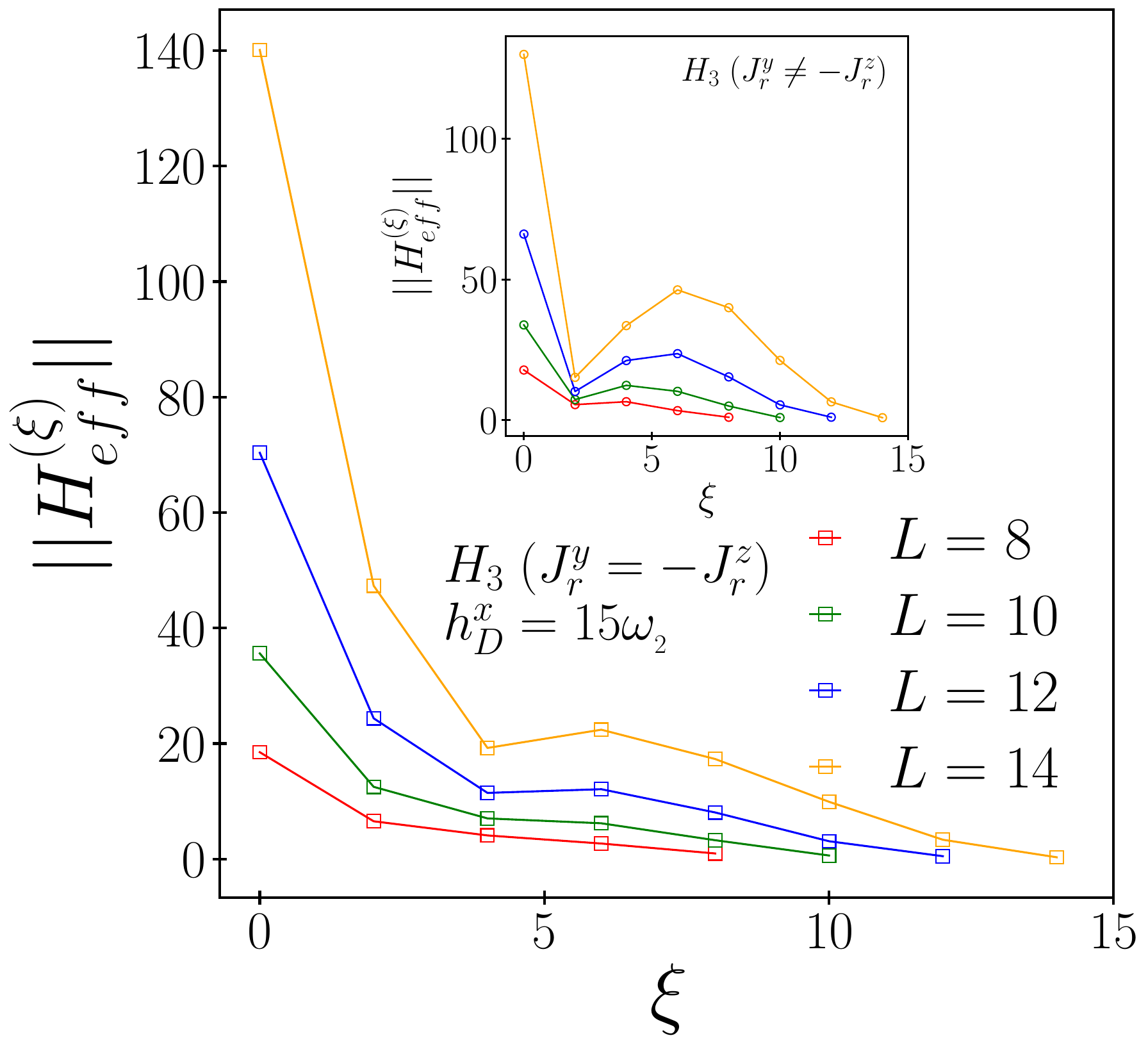}
\hfill
\includegraphics[width=0.47\textwidth]
{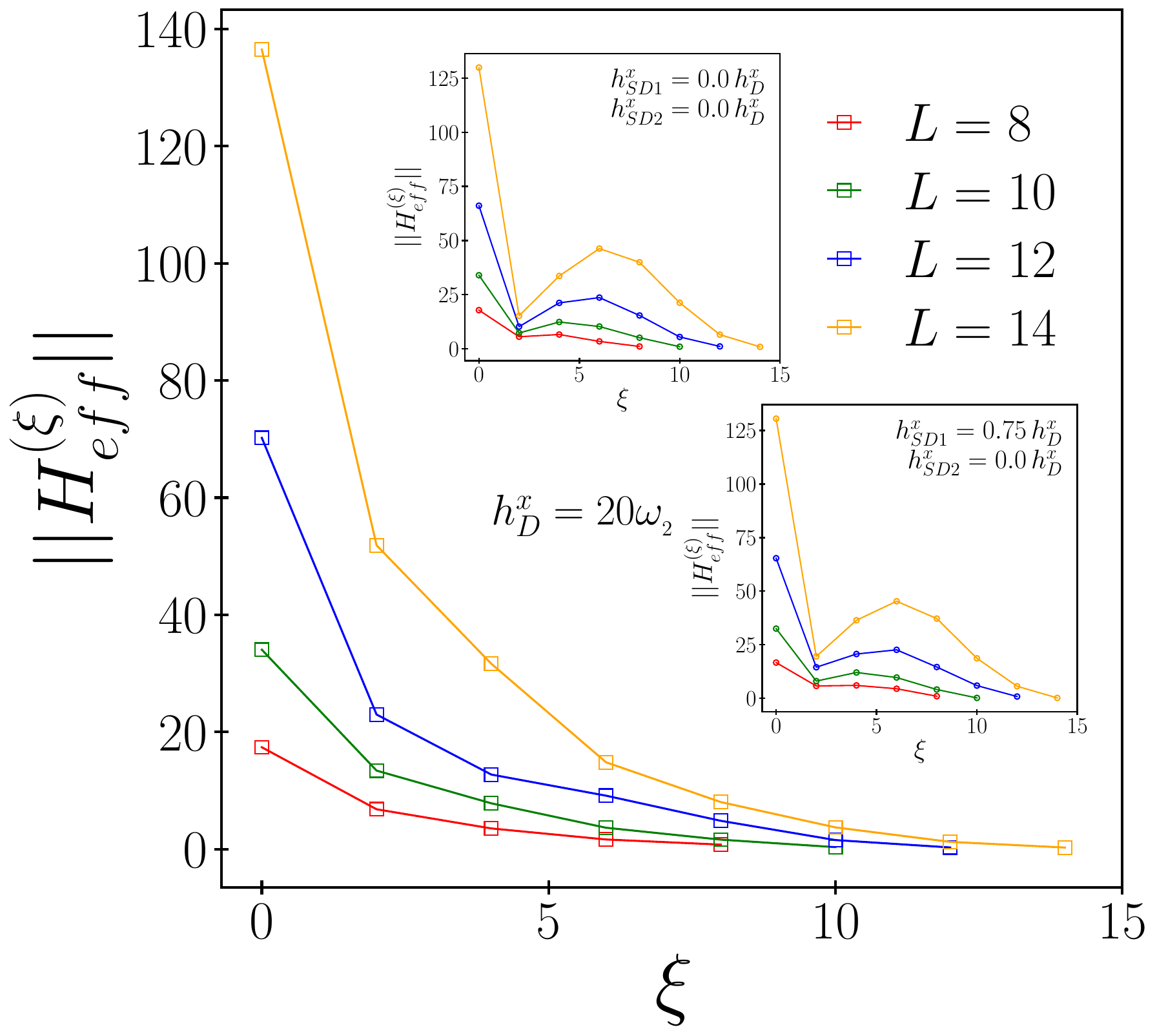}

\caption{
\textbf{Spin-flip ordered series of the effective Floquet Hamiltonian.}
The Frobenius norm $\|H_{\mathrm{eff}}^{(\xi)}\|$ is shown as a
function of the Hamming distance $\xi$ between $x$-basis
configurations for $L=8,10,12,14$.
\textbf{Top left:} For the field-driven Ising model $H_1(t)$, the
weak-drive case $h_D^x=2\omega_{_1}$ exhibits substantial weight at
higher spin-flip orders, whereas for the strong drive
$h_D^x=30\omega_{_1}$ the weight is concentrated in the low-$\xi$
sectors and rapidly decreases with $\xi$. For the strong-drive case,
the decay sets in at approximately the same spin-flip order for all
system sizes.
\textbf{Top right:} For $H_2(t)$, the combined bond drive with
$J_D=40\omega_{_1}$ and $J_{SD}=25\omega_{_1}$ produces a rapidly
decaying spin-flip ordered profile. The insets show the corresponding
profiles for the uniform drive, $J_D=40\omega_{_1}$, $J_{SD}=0$, and
the staggered drive, $J_D=0$, $J_{SD}=40\omega_{_1}$, which retain
substantial weight at higher spin-flip orders.
\textbf{Bottom left:} For $H_3(t)$, time-independent Hamiltonian
engineering with $J_r^y=-J_r^z$ and $h_D^x=15\omega_{_2}$ yields a
spin-flip ordered weight that decreases systematically with $\xi$,
whereas the unengineered case $J_r^y\neq -J_r^z$ retains appreciable
weight at larger $\xi$.
\textbf{Bottom right:} For $H_3(t)$ with $h_D^x=20\omega_{_2}$, the
fully modulated drive with $h_{SD,1}^x=0.75h_D^x$ and
$h_{SD,2}^x=0.5h_D^x$ gives a rapidly decaying profile. The insets
compare the unmodulated drive, $h_{SD,1}^x=h_{SD,2}^x=0$, and the
partially modulated drive, $h_{SD,1}^x=0.75h_D^x$,
$h_{SD,2}^x=0$. Across the engineered cases, the large-$\xi$ weight is
strongly suppressed, consistent with the quasi-local structure of the
effective Floquet Hamiltonian.
}
\label{fig:SM_LOE_engineering}
\end{figure*}

We next examine whether a similar suppression of large-$\xi$ processes
can be obtained in the models for which dynamical freezing is not stable
without engineering. For $H_2(t)$, the drive-engineering protocol combines
uniform and staggered bond driving. The top-right panel of
Fig.~\ref{fig:SM_LOE_engineering} shows the spin-flip ordered profile for
$J_D=40\omega_{_1}$ and $J_{SD}=25\omega_{_1}$. The Frobenius norm rapidly
decreases with increasing $\xi$ for all system sizes considered. The upper
and lower insets show the corresponding profiles for the uniform drive,
$J_D=40\omega_{_1}$, $J_{SD}=0$, and the staggered drive,
$J_D=0$, $J_{SD}=40\omega_{_1}$. In these two cases, appreciable weight
remains at larger spin-flip orders, demonstrating that the combined drive
is required to suppress the large-$\xi$ contributions.\\

For $H_3(t)$, we consider two complementary engineering routes. The first
is time-independent Hamiltonian engineering. For $J_r^y=-J_r^z$ and
$h_D^x=15\omega_{_2}$, the spin-flip ordered weight decreases
systematically with $\xi$ for all system sizes, as shown in the bottom-left
panel of Fig.~\ref{fig:SM_LOE_engineering}. In contrast, for the
unengineered Hamiltonian with $J_r^y\neq -J_r^z$, substantial weight
persists at large spin-flip distances. The second route is drive
engineering. For $h_D^x=20\omega_{_2}$, the fully modulated drive with
$h_{SD,1}^x=0.75h_D^x$ and $h_{SD,2}^x=0.5h_D^x$ produces a rapidly
decaying spin-flip ordered profile. The insets compare this result with
the unmodulated drive, $h_{SD,1}^x=h_{SD,2}^x=0$, and the partially
modulated drive, $h_{SD,1}^x=0.75h_D^x$, $h_{SD,2}^x=0$. Both incomplete
drive modulations retain substantially more weight at larger $\xi$.\\

Taken together, these results show that the spin-flip ordered series distinguishes the dynamically frozen regime from the unstable strongly driven regime. Strong driving alone does not guarantee a rapidly decaying spin-flip ordered profile, as illustrated by the un-engineered $H_2(t)$ and $H_3(t)$ cases. Instead, the engineering protocols identified from the rotating-frame criterion suppress large-$\xi$ processes and produce an effective Hamiltonian whose
spin-flip ordered weight is concentrated within a system-size-independent range of $\xi$, consistent with the quasi-local description discussed in the main text.

\end{document}